\documentclass[twocolumn,a4paper,notitlepage]{article}
\usepackage{fullpage}
\usepackage[affil-it]{authblk}
\usepackage{graphicx}
\usepackage[pdftex]{hyperref}
\usepackage{subfig}
\usepackage{bm}
\usepackage{amsmath, amsthm, amssymb}
\usepackage{url}

\newcommand{\kk}{\mathbf{k}}
\newcommand{\eqn}[1]{\begin{equation} #1\end{equation}}
\newcommand{\iprod}[2]{\left< #1 \right| \left. #2 \right>}

\begin{document}
\title{Wavelet methods to eliminate resonances \\
in the Galerkin-truncated Burgers and Euler equations}

\author{R. M. Pereira}
\affil{Instituto de F\'{i}sica, Universidade Federal do Rio de Janeiro, CP 68528, 21945-970, Rio de Janeiro, RJ, Brazil}
\affil{Divis\~{a}ão de Metrologia em Din\^{a}mica de Fluidos, Instituto Nacional de Metrologia, Normaliza\c{c}\~{a}o e Qualidade Industrial, Av. Nossa Senhora das Gra\c{c}as 50, Duque de Caxias, 25250-020, Rio de Janeiro, Brazil}
\affil{LMD-CNRS-IPSL, \'Ecole Normale Sup\'erieure, 24 rue Lhomond, 75231 Paris Cedex 5, France}
\author{R. Nguyen van yen}
\affil{Fachbereich Mathematik und Informatik, Freie Universit\"at Berlin, Arnimallee 6, D-14195 Berlin, Germany}
\affil{LMD-CNRS-IPSL, \'Ecole Normale Sup\'erieure, 24 rue Lhomond, 75231 Paris Cedex 5, France}
\author{M. Farge}
\affil{LMD-CNRS-IPSL, \'Ecole Normale Sup\'erieure, 24 rue Lhomond, 75231 Paris Cedex 5, France}

\author{K. Schneider}
\affil{M2P2-CNRS and CMI, Universit\'e d'Aix-Marseille, 38 rue F. Joliot-Curie, 13451 Marseille Cedex 20, France}

%
\maketitle

\begin{abstract} 
It is well known that solutions to the Fourier-Galerkin truncation of the inviscid Burgers equation (and other hyperbolic conservation laws)
do not converge to the physically relevant entropy solution after the formation of the first shock.
This loss of convergence was recently studied in detail in [S. S. Ray \textit{et al.}, \textit{Phys. Rev. E} \textbf{84}, 016301 (2011)],
and traced back to the appearance of a spatially localized resonance phenomenon perturbing the solution.
In this work, we propose a way to remove this resonance by filtering a wavelet representation of the Galerkin-truncated equations.
A method previously developed with a complex-valued wavelet frame is applied and expanded to embrace the use of real-valued orthogonal wavelet basis, 
which we show to yield satisfactory results only under the condition of adding a safety zone in wavelet space. 
We also apply the complex-valued wavelet based method to the 2D Euler equation problem, showing that it is able to filter the resonances in this case as well.
\end{abstract}


\section{Introduction}

Due to the intrinsic limitations of computers, solving a nonlinear partial differential equation numerically actually means solving its truncation to a finite number of modes,
where, in favorable cases, the truncated system closely approaches its continuous counterpart.
But sometimes the truncation has drastic effects which completely destroy the desired approximation. 
The first historical example for which this happened was probably the symmetric finite difference scheme designed by von Neumann in the 1940s for nonlinear conservation laws.
As recalled in \cite{Hou1991}, it was indeed shown in the 1980s that, when applying this scheme even to the simplest case of the 1D inviscid Burgers equations, 
convergence to the correct solution is lost at the appearance of the first shock.
Other schemes, specifically designed to dissipate kinetic energy at the location of shocks, do not suffer from this limitation and yield the desired solution.

This matter of convergence was investigated in \cite{Tadmor1989} for another important scheme, namely Fourier-Galerkin truncation,
where only the equations for Fourier modes with wavenumbers below a certain cut-off are solved, the other modes being set to zero.
Using the conservative character of the truncation and the nonlinear structure of the equations, the author was able to prove that even weak convergence to the physical solutions was ruled out once the latter started to be dissipative.
This loss of convergence was scrutinized more closely in the recent work \cite{Ray2011}, which showed that in the truncated system
shocks become sources of waves that perturb the numerical solution throughout its spatial domain. 
This is possible because Fourier-Galerkin truncation is a non-local operator in physical space, instantaneously removing all modes above the truncation wavenumber.
Furthermore, these waves resonantly interact with the flow at locations where the velocity is the same as their phase velocity, giving rise to strong perturbations localized around these positions which eventually spread and corrupt the numerical solution. 

The aim of the present work is to show how the resonances can be eliminated by filtering the solution in a wavelet basis, a possibility which was already pointed out in \cite{Ray2011}. 
The Burgers equation has been chosen as a toy model because its entropy solutions can be computed analytically, enabling direct comparison with numerical results. 
An important point to keep in mind though is that the analytical solutions are dissipative even in the inviscid limit, 
a phenomenon known as dissipative anomaly, while the Galerkin-truncated ones never dissipate energy if the viscosity is set to zero. 
Therefore a numerical solution can approach the exact solution only if it finds a way to dissipate energy, as is achieved by our method through the filtering process
described further down.
In fact, as discussed in \cite{Nguyenvanyen2008,Nguyenvanyen2009} and references therein, 
many filtering mechanisms are known empirically to achieve this task (see also the recent review in \cite{Gottlieb2011}).
However, the precise effect of these filtering methods on the resonances shown by \cite{Ray2011} has not been fully clarified yet.

To get insight into the formation of the resonance we start by performing a continuous wavelet analysis of the Galerkin-truncated solutions to the inviscid Burgers equation. 
Such a representation unfolds the solution in both space and scale in a continuous fashion. 
It thus allows to visualize at which wavenumbers and positions the resonances are generated and subsequently propagated.

Afterwards, the wavelet filtering method analogous to Coherent Vorticity Simulation (CVS), 
already proposed to solve Burgers equation \cite{Nguyenvanyen2008,Nguyenvanyen2009}, is applied here with the same initial conditions used in \cite{Ray2011}.
To demonstrate that the method is well suited for regularizing the solution, the equation is solved in Fourier space using a pseudo-spectral approach, but after each time step the solution is expanded over a frame of complex-valued wavelets, filtered with an iterative procedure introduced in \cite{AAMF04}, and then reprojected onto the Fourier basis for computing the next time step. 

We then go further and propose the use of real-valued orthogonal wavelets instead of the redundant complex-valued wavelets. 
Since the former do not enjoy the translational invariance property of the latter, satisfactory solutions can only be obtained by keeping the neighbors of the retained coefficients, \emph{i.e.}, adding a safety zone in wavelet coefficient space to account for the shocks translation and the small scale generation, a procedure successfully applied in previous works for 2D and 3D flows \cite{Froehlich1999, Schneider2006, Okamoto2011}. 
The quality of the approximations obtained for the different filtering methods is assessed by computing a global error estimate.

Finally, since \cite{Ray2011} also discusses the presence of resonances in the Galerkin-truncated 2D incompressible Euler equations,
we accordingly study the effect of the complex-valued wavelet method in this case. 
First results in that same direction can be found in \cite{Nguyenvanyen2009}.


\section{1D inviscid Burgers equation}

\subsection{Continuous wavelet analysis}\label{seq:CWT}

Our starting point is the inviscid Burgers equation, written in conservative form
\eqn{ \partial_t u +\frac{1}{2} \partial_x u^2 = 0, \label{burgers}}
$u$ being velocity, $t$ time and $x$ space, plus periodic boundary conditions, and taking the same harmonic initial condition as in \cite{Ray2011} (the domain size being normalized to 1):
\eqn{u_0(x) = \sin(2\pi x) + \sin(4\pi x + 0.9) + \sin(6\pi x). \label{init_cond}}
In \cite{Ray2011} the authors observed that, when solving the Galerkin-truncated version of (\ref{burgers}) with a pseudo-spectral code, fine scale oscillations appear all over the solution right after the formation of the first singularity in the exact solution, followed by the emergence of two bulges around the points having the shock velocity with positive velocity gradient. 
These bulges then grow and start to perturb the solution, initiating the equipartition process predicted by T.D. Lee \cite{Lee1952}. 
As explained in \cite{Ray2011}, the bulges are due to a resonant interaction between a truncation wave, excited by the Gibbs oscillations coming from the Galerkin truncation, and the locations where the velocities are close to the phase velocity of the wave.

\begin{figure*}
\centering
\subfloat{\label{fig:cwt_0}\includegraphics[width=0.45\textwidth]{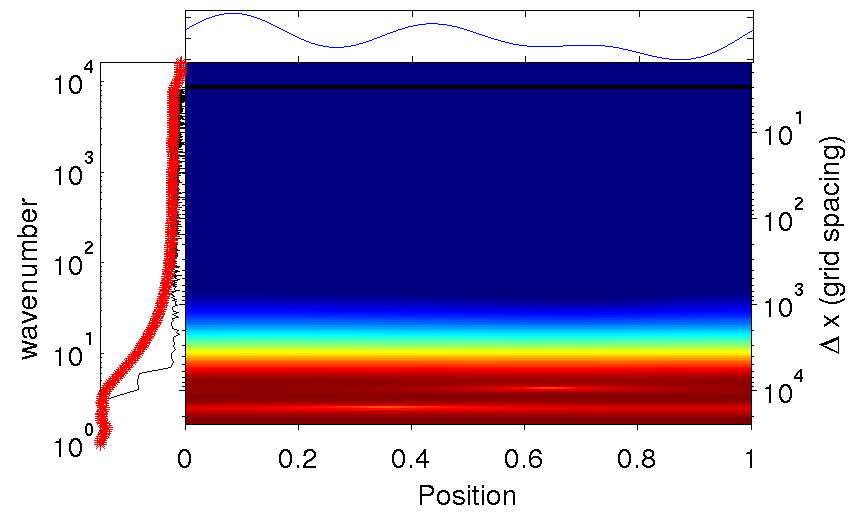}}\hspace{0.5cm}
\subfloat{\label{fig:cwt_02749}\includegraphics[width=0.45\textwidth]{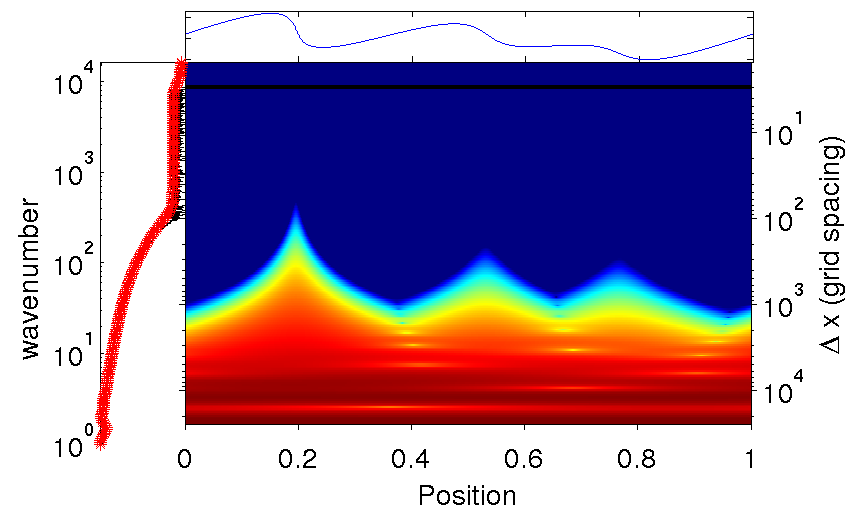}} \\
\subfloat{\label{fig:cwt_03505}\includegraphics[width=0.45\textwidth]{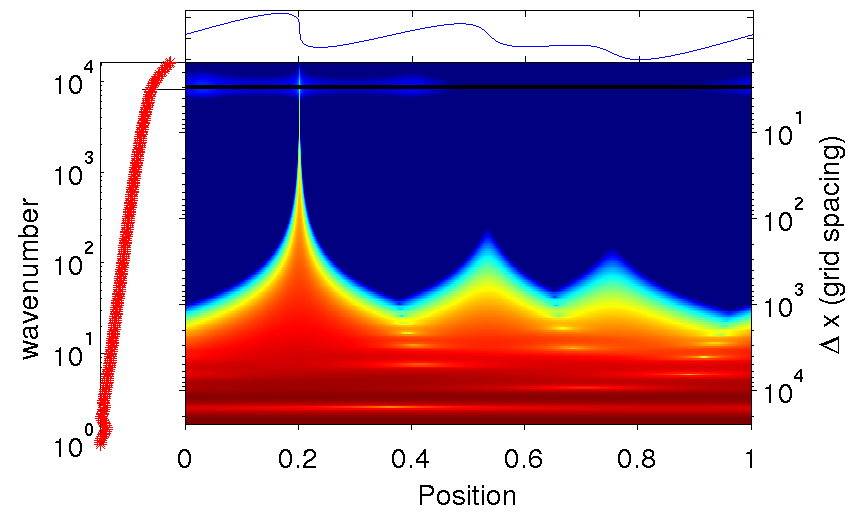}} \hspace{0.5cm}
\subfloat{\label{fig:cwt_03538}\includegraphics[width=0.45\textwidth]{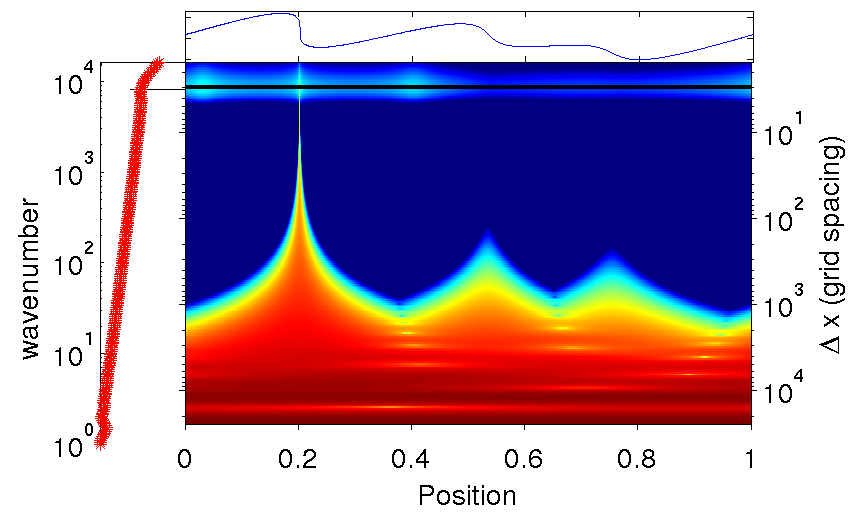}} \\
\subfloat{\label{fig:cwt_03648}\includegraphics[width=0.45\textwidth]{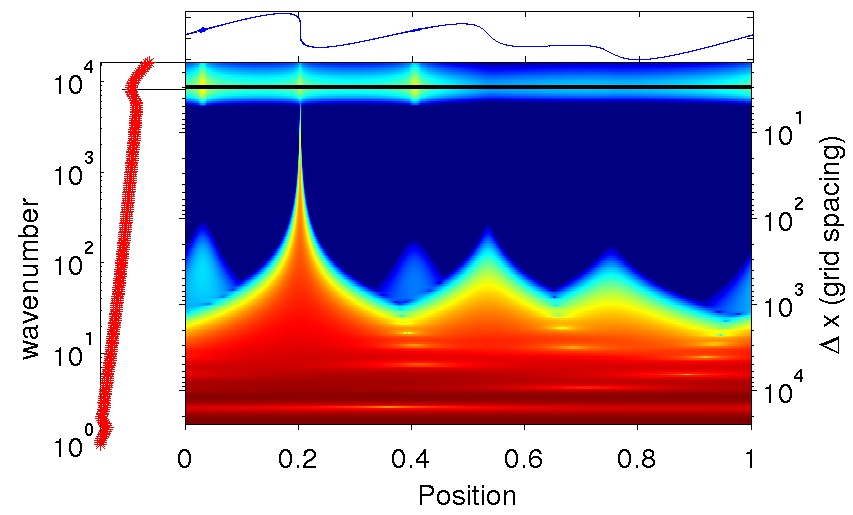}} \hspace{0.5cm}
\subfloat{\label{fig:cwt_03998}\includegraphics[width=0.45\textwidth]{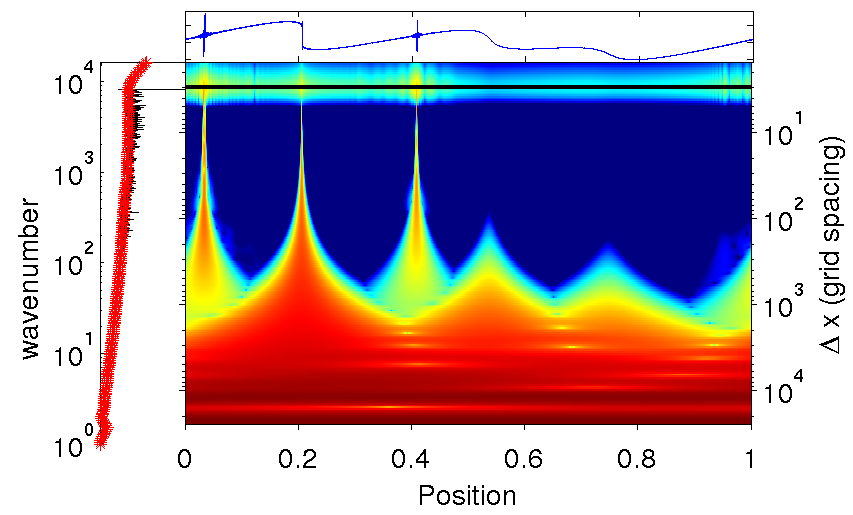}} \\ 
\subfloat{\label{fig:cwt_05897}\includegraphics[width=0.45\textwidth]{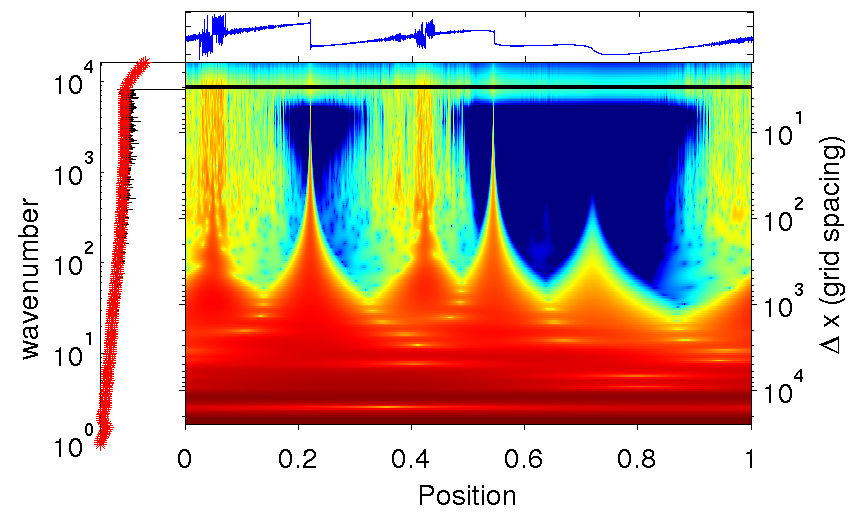}} \hspace{0.5cm} 
\subfloat{\label{fig:cwt_19989}\includegraphics[width=0.45\textwidth]{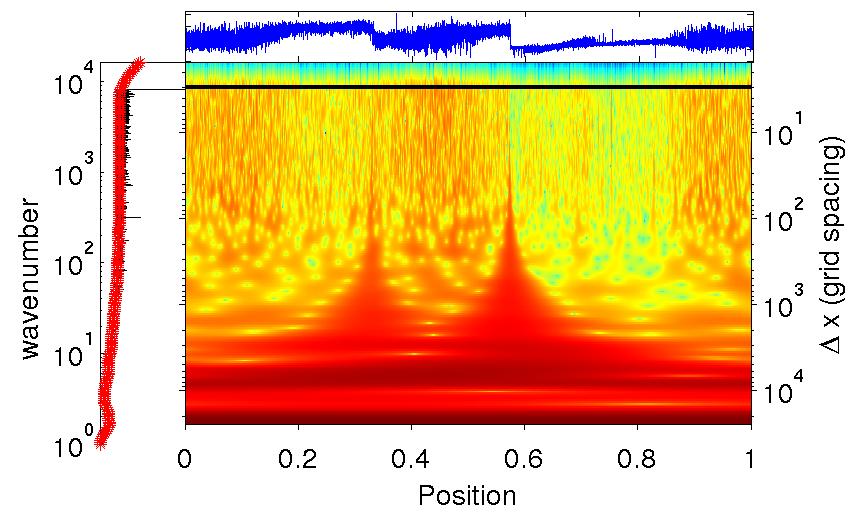}} \\ 
\subfloat{\includegraphics[width=0.5\textwidth]{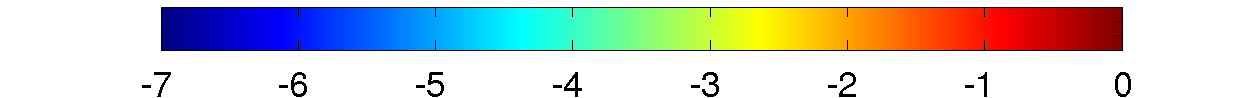}}
\caption{
Base 10 logarithm of moduli of CWT coefficients of the Galerkin-truncated solution, at time 
$t=0$ (a), $t=0.02749$ (b), $t=0.03505$ (c), $t=0.03538$ (d), $t=0.03648$ (e), $t=0.03998$ (f), $t=0.05897$ (g), and $t=0.19989$ (h).
Each subfigure shows the solution (on the top) and below the base 10 logarithm of the modulus of the corresponding CWT coefficients. 
The corresponding wavenumber spectrum is plotted vertically on the left. 
}
\label{fig:cwt}
\end{figure*}

To follow the formation of resonances and the subsequent spreading of the fluctuations, let us first consider the continuous wavelet transform (CWT) of the numerical solution at different time instants. 
All computations were performed using a 4$^\mathrm{th}$ order Runge-Kutta time evolution scheme with $\delta t = 0.125 N^{-1}$ as time step, up to a Galerkin truncation wavenumber $K = 8192$.
For efficiency, the nonlinear term is computed pseudo-spectrally on a collocation grid having $3K$ points, which ensures full dealiasing.
The CWT coefficients are calculated as the inner products of the velocity $u(x)$ at a given instant $t$ with a set of wavelet functions $\psi_{l,x'}(x)$ of scales $\ell$ centered around positions $x'$,
where for the mother wavelet we have chosen the complex-valued Morlet wavelet for its good analysis properties \cite{Farge92}.
The results, presented in Fig.~\ref{fig:cwt}, show the logarithm of the modulus of wavelet coefficients at different positions $x'$ and scales $\ell$ 
(represented by the equivalent wavenumbers $k =\frac{k_\psi}{\ell}$, $k_\psi$ being the centroid wavenumber of the chosen wavelet \cite{Ruppert-Felsot2009}). 
The horizontal black line indicates the Galerkin truncation frequency and the velocity fields themselves are also shown at the top of each figure for convenience.

Figures~\ref{fig:cwt_0} and \ref{fig:cwt_02749} show respectively the harmonic initial condition and how the precursors of the shocks develop. 
Figure \ref{fig:cwt_03505} shows the solution when the first preshock reaches the cut-off scale and becomes a shock, \emph{i.e.}, when non negligible energy reaches the scale indicated by the horizontal black line. 
We observe that the first resonances appear immediately after that 
(note the small time interval between Figs.~\ref{fig:cwt_03505} and~ \ref{fig:cwt_03538}) 
and then spread all over space. 
Figure~\ref{fig:cwt_03648} shows the formation of the bulges around the resonant locations. 
They stretch until they reach the Galerkin scale and then generate more truncation waves, as shown in Fig.~\ref{fig:cwt_03998}. 
After that, perturbations at all scales start to spread throughout the solution, and even more so when the second shock is formed, as in Fig.~\ref{fig:cwt_05897}. 
For much longer time the solution then becomes very noisy (Fig.~\ref{fig:cwt_19989}), on its way towards equipartition\footnote{Videos with the time evolution of the coefficients were made available on-line for the interested reader as supplementary material to this paper, and also at \url{http://www.youtube.com/watch?v=WX2YIHGR7LA} and \url{http://www.youtube.com/watch?v=j4VfBGgSy30}.}.


\subsection{Elimination of resonances using complex-valued Kingslets}\label{sec:kingslet}

As explained in \cite{Ray2011}, and as we have seen from the wavelet analysis of the previous section, 
the failure of the Fourier-Galerkin scheme to reproduce the correct solution can be traced back to the amplification of truncation waves by a resonance mechanism.
To suppress these resonances, a dissipation mechanism has to be introduced in the numerical scheme,
in a way which does not affect the nonlinear dynamics.
This procedure is sometimes called regularization of the solution.
In this section, we show by numerical experiments how the resonances are suppressed by the CVS-filtering method, which was first applied to the inviscid Burgers equation in \cite{Nguyenvanyen2008},
and recall its interpretation in terms of denoising.

The algorithm proposed by \cite{Nguyenvanyen2008} is as follows.
Starting from the Fourier coefficients of the velocity field $\widehat{u}_k$ for $\vert k \vert  \leq K$ at $t=t_n$: 
\begin{enumerate}
\item \emph{Time integration}. 
The Fourier coefficients of the velocity field are advanced in time to $t=t_{n+1}$ using the $4^{th}$ order Runge-Kutta scheme described in Sec.~\ref{seq:CWT}.

\item \emph{Inverse Fourier transform}. The velocity field at $t=t_{n+1}$ is reconstructed from its Fourier coefficients on a grid with $N=2K$ points.

\item \emph{Forward wavelet transform}. The velocity field is written in wavelet space as
\eqn{ u(x) = \iprod{\phi}{u} \phi(x) + \sum_{j=0}^{J-1}\sum_{i=1}^{2^j} \iprod{\psi_{ji}}{u} \psi_{ji}(x), \label{wavelet_expansion}}
where $\psi_{ji}$ are the wavelet functions, $\phi$ the associated scaling function and the indexes $j$ and $i$ denote scale and position respectively. Each inner product, defined as $\iprod{f}{g} \equiv \int_0^1 f(x)^*g(x)dx$, corresponds to a wavelet coefficient.

\item \emph{Application of the CVS filter}. The coefficients whose modulus are below a threshold $T$, so-called incoherent coefficients, are discarded, and $T$ is determined at each time step in an iterative way following \cite{AAMF04}. It is initialized as $T_0 = q\sqrt{E/N}$, $q$ being a compression parameter and $E$ being the total energy, then successive filterings are made as $T$ is recalculated in sub-step $n+1$ as
\eqn{ T_{n+1} = q\, \sigma\!\left[ \tilde u^{(n)}_{ji} \right] \label{threshold},}
until $T_{n+1}=T_n$. Here $\tilde u_{ji}^{(n)}$ are the wavelet coefficients below the threshold $T_n$ and $\sigma[\cdot]$ represents the standard deviation of the set of coefficients between brackets. 

\item \emph{Inverse wavelet transform}.
The coefficients above the final threshold represent the coherent part of the signal and are used as input to an inverse fast wavelet transform.

\item \emph{Forward Fourier transform}.
The Fourier coefficients of the filtered velocity field are computed, and the cycle can proceed onward.
\end{enumerate}

There are two choices left to be made in this algorithm: the wavelet basis used in steps 3 and 5, and the parameter $q$ in step 4.
As shown in \cite{Nguyenvanyen2009}, this version of the algorithm performs badly if real-valued orthogonal wavelets are used,
but works very well when using translation invariant complex-valued wavelets called Kingslets, introduced in \cite{NK01}
and first proposed in \cite{Nguyenvanyen2008} for this application.
Note that Kingslets were constructed to have almost vanishing energy in the negative wavenumber range,
which (as explained in \cite{NK01}) implies that filtering in wavelet space is almost a translation invariant operator (i.e., it commutes with spatial translations of the signal).
This is a desired feature for Burgers equation since shocks translate and cannot be properly tracked with a real-valued wavelet basis, 
whose coefficients are not stable enough due to the loss of translational invariance, giving poor filtering results.
Therefore, we stick to this choice in this section, but we will discuss below how the algorithm can be modified to authorize other choices.

Concerning the dimensionless number $q$ in step 4 of the algorithm,
it controls the severity of the filter, since increasing $q$ enlarges the set of discarded coefficients.
Its value defines a certain balance between regularization and approximation quality, and also influences the compression rate.
Here, we follow \cite{Nguyenvanyen2008} and use $q=5$ with Kingslets. 
A discussion of the effect of varying $q$ would be of interest but is out of the scope of the present work.
 
The added complexity of running this algorithm, as compared to the standard Fourier-Galerkin method, comes from 
the forward and inverse Fourier and wavelet transforms, and the iterations required to determine the threshold.
Since the standard 4-th order Runge-Kutta scheme already requires $12$ Fourier transforms per timestep,
the additional Fourier transforms represent an increase of computational cost of about 17\% in total.
The cost of each wavelet transform is proportional to $S\log_2(N)$ where $S$ is the length of the wavelet filters, 
and for efficient implementations it is lower than the cost of a Fourier transform.
Finally, the cost of the iterations is more difficult to evaluate since their number is not known a priori, 
but we observe in practice that it is low compared to the other costs.


%
\begin{figure*}
\centering
\subfloat{\label{fig:kings_nofil_037}\includegraphics[width=0.32\textwidth, trim=0cm 0cm 0.8cm 0cm, clip=true]{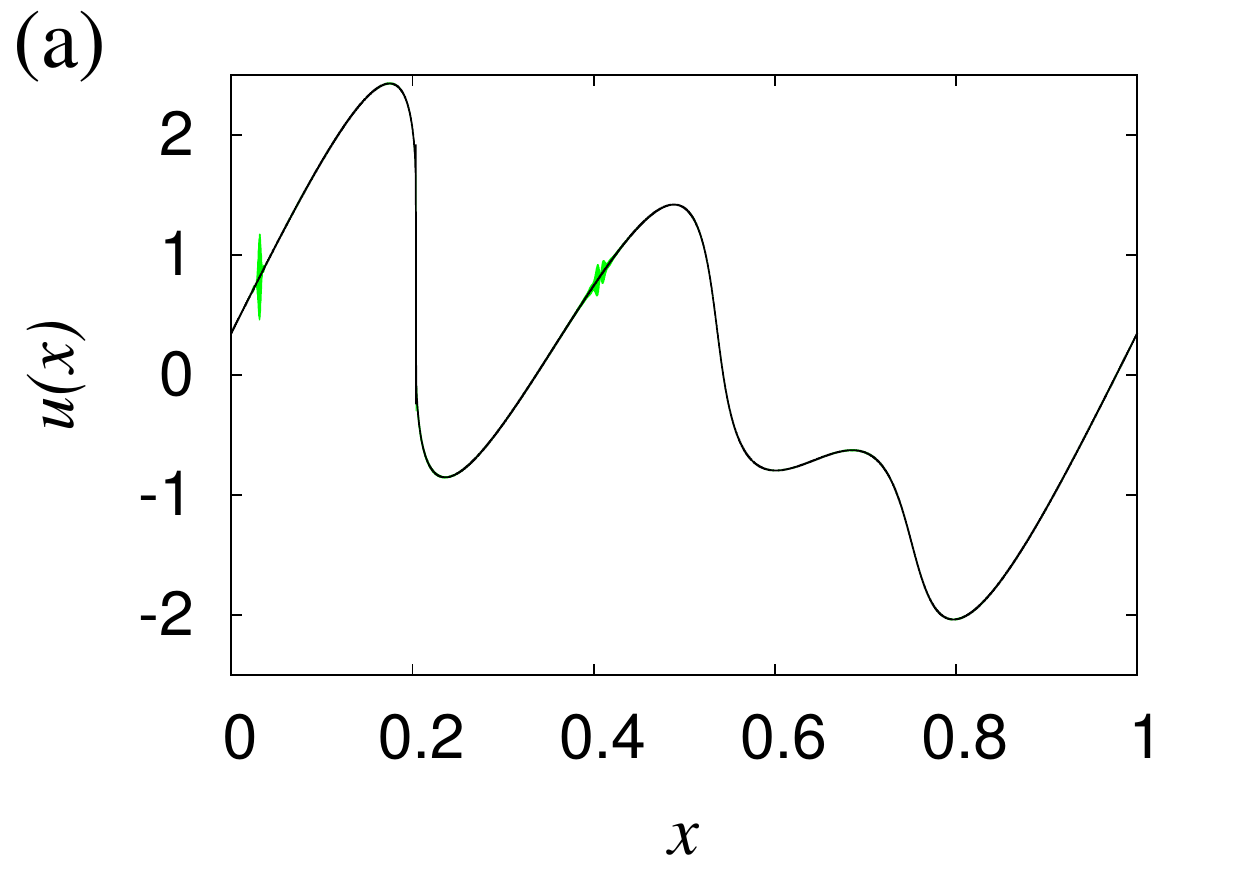}}\hspace{0.15cm}
\subfloat{\label{fig:kings_nofil_048}\includegraphics[width=0.32\textwidth, trim=0cm 0cm 0.8cm 0cm, clip=true]{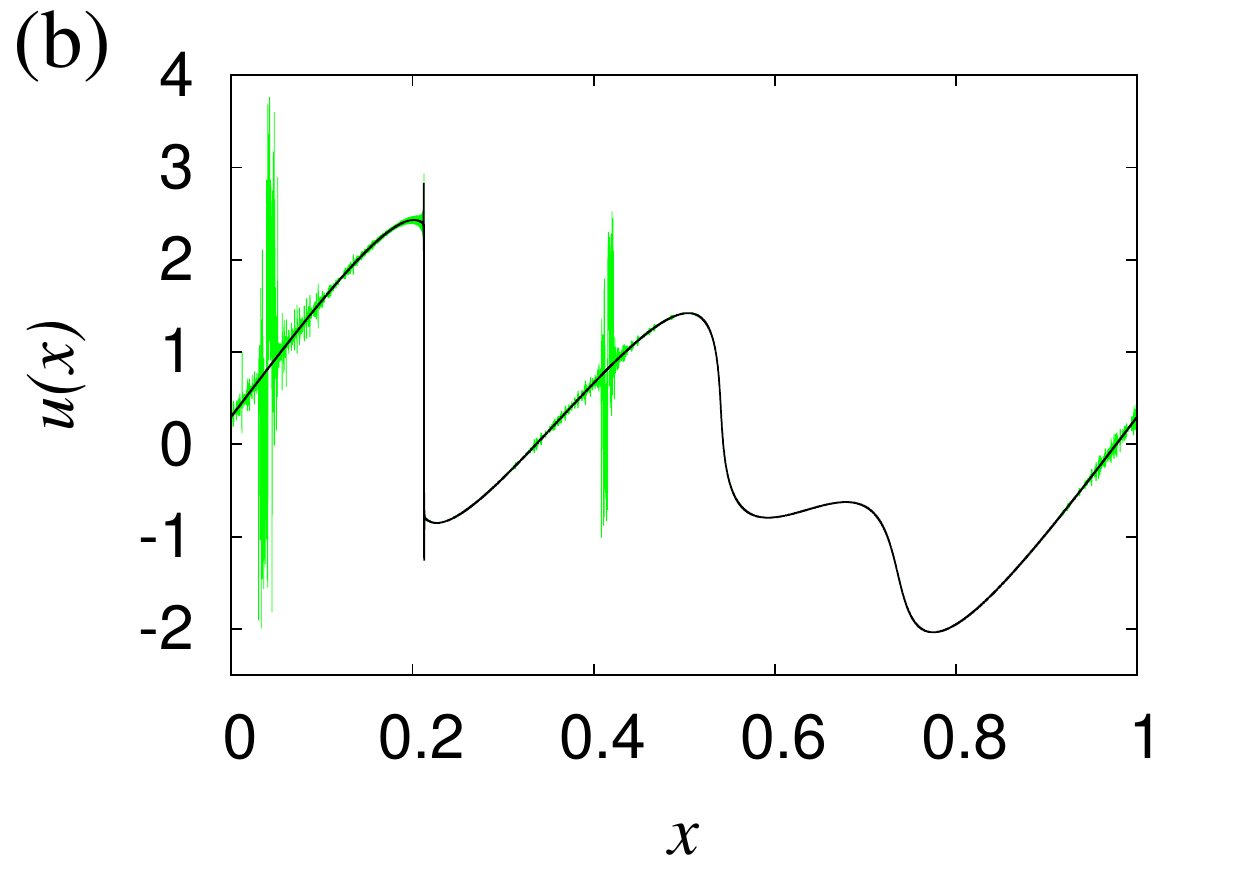}}\hspace{0.15cm}
\subfloat{\label{fig:kings_nofil_129}\includegraphics[width=0.32\textwidth, trim=0cm 0cm 0.8cm 0cm, clip=true]{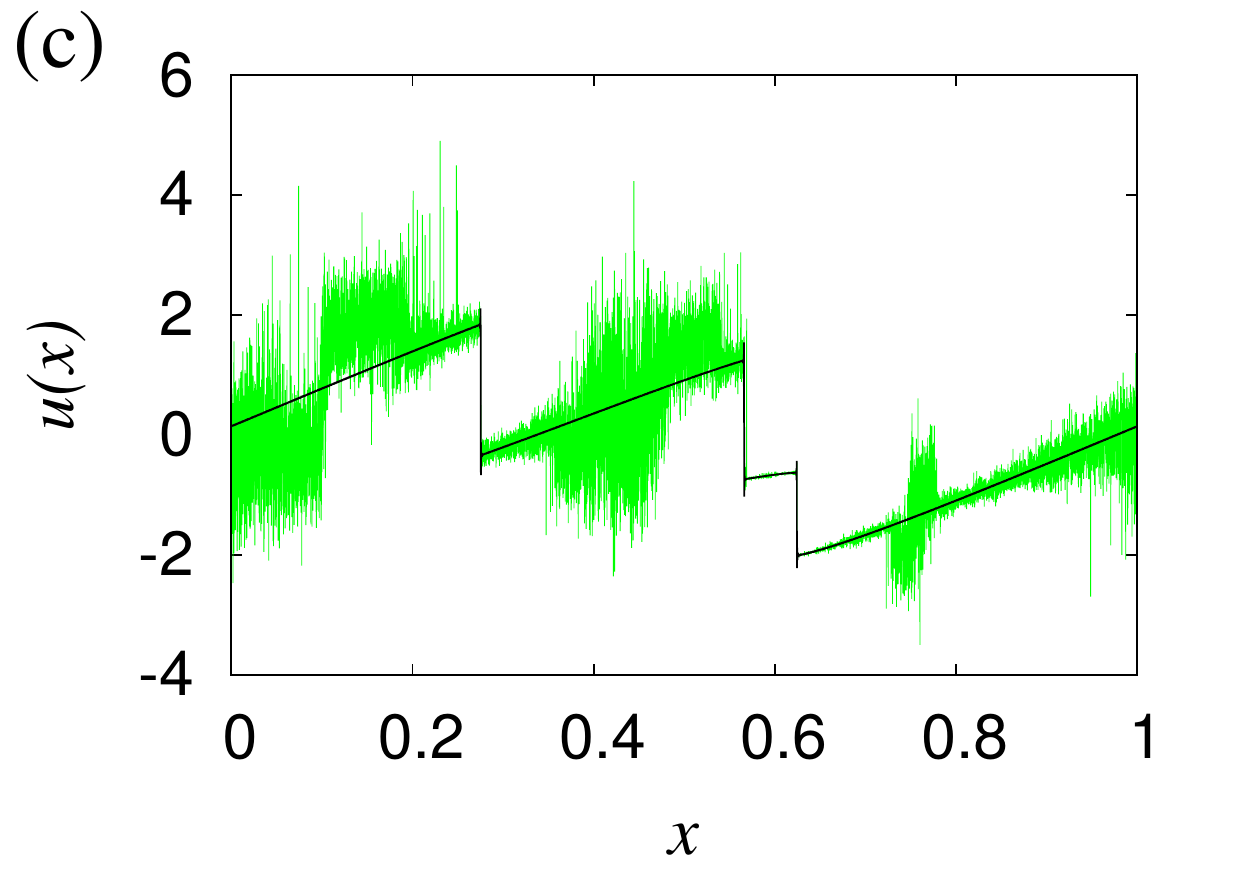}}
\caption{ 
Solutions of the truncated inviscid Burgers equation at $t=0.037$ (a), $t = 0.048$ (b), and $t = 0.129$ (c).
Green (gray): Galerkin-truncated solutions. Black: CVS-filtered with Kingslets.}
\label{fig:kings_nofil}
\end{figure*}
\begin{figure}
\centering
\includegraphics[width=0.4\textwidth]{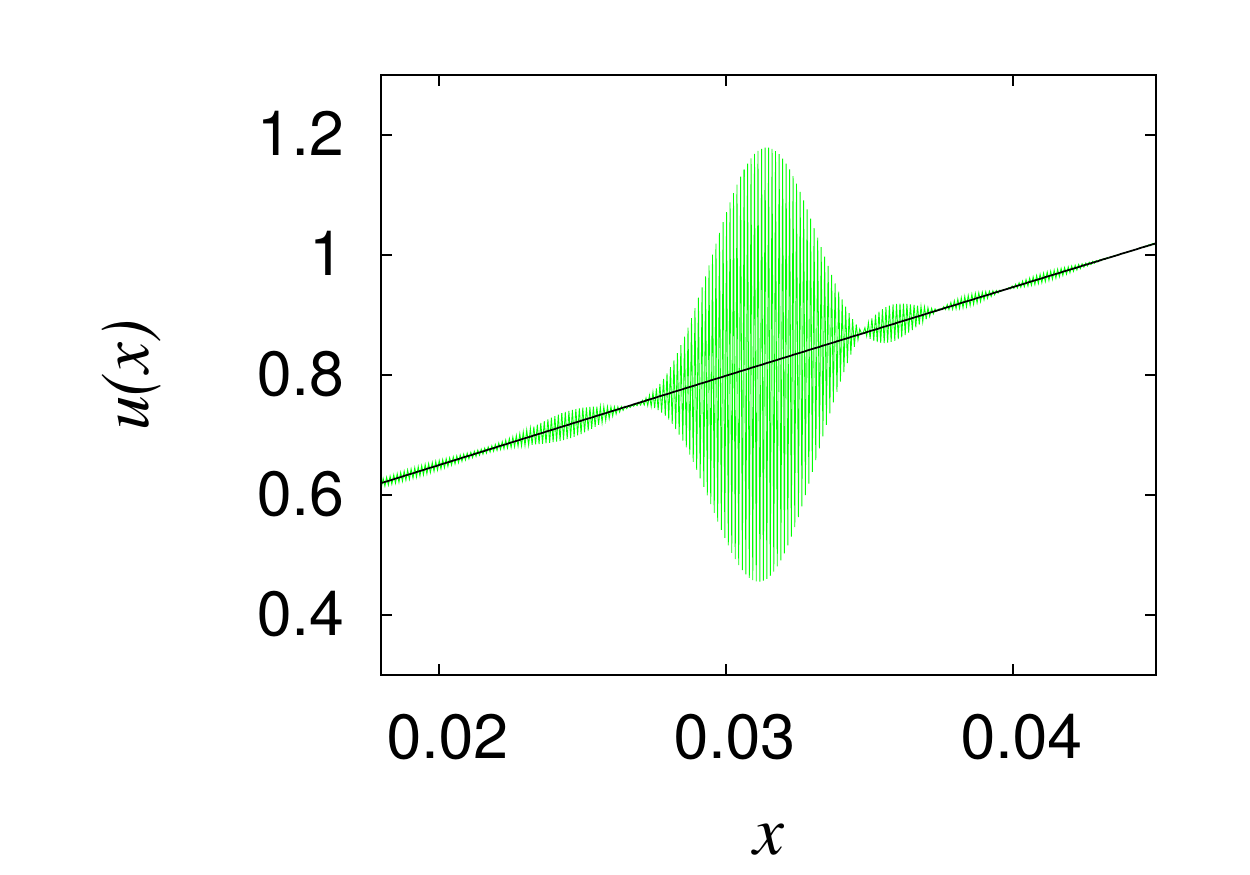}
\caption{ Zoom of resonance at $t = 0.037$. Green (gray): Galerkin-truncated solutions. Black: CVS-filtered with Kingslets.}
\label{fig:kings_tyg}
\end{figure}

In Fig.~\ref{fig:kings_nofil_037} we show the solutions a few time steps after the appearance of the resonances, which do not occur for the CVS-filtered solution (shown in black). 
Figures~\ref{fig:kings_nofil_048} and~\ref{fig:kings_nofil_129} show that the evolution is stable and we still have no trace of resonances, even for longer integration times when the Galerkin-truncated solution becomes perturbed, 
although after the formation of shocks the Gibbs phenomenon is intense (as discussed in \cite{Nguyenvanyen2008, Nguyenvanyen2009}). 
In Fig.~\ref{fig:kings_tyg} we show in detail how the resonances are completely filtered out by the CVS method.

\begin{figure*}
\centering
\subfloat{\label{fig:kings_ref_037}\includegraphics[width=0.32\textwidth, trim=0cm 0cm 0.8cm 0cm, clip=true]{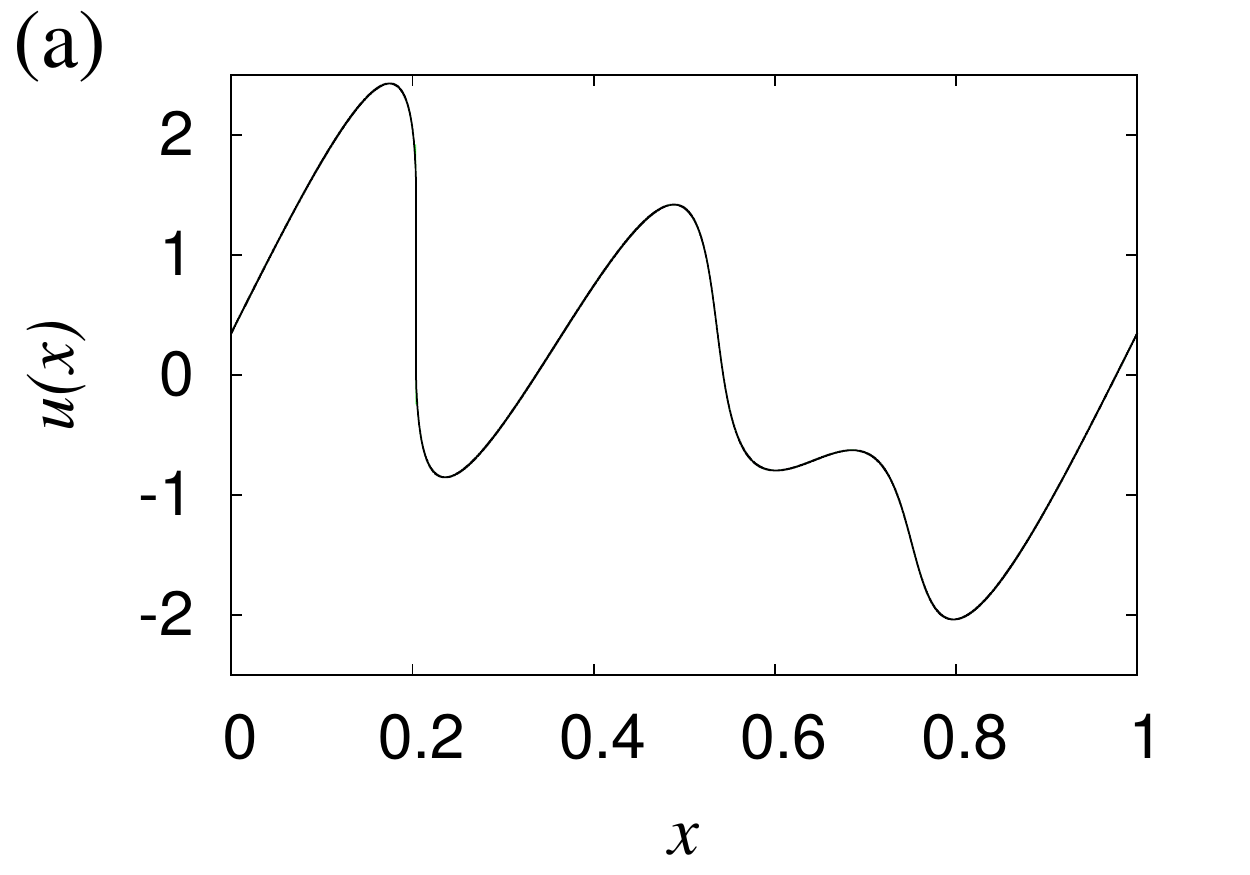}}\hspace{0.15cm}
\subfloat{\label{fig:kings_ref_048}\includegraphics[width=0.32\textwidth, trim=0cm 0cm 0.8cm 0cm, clip=true]{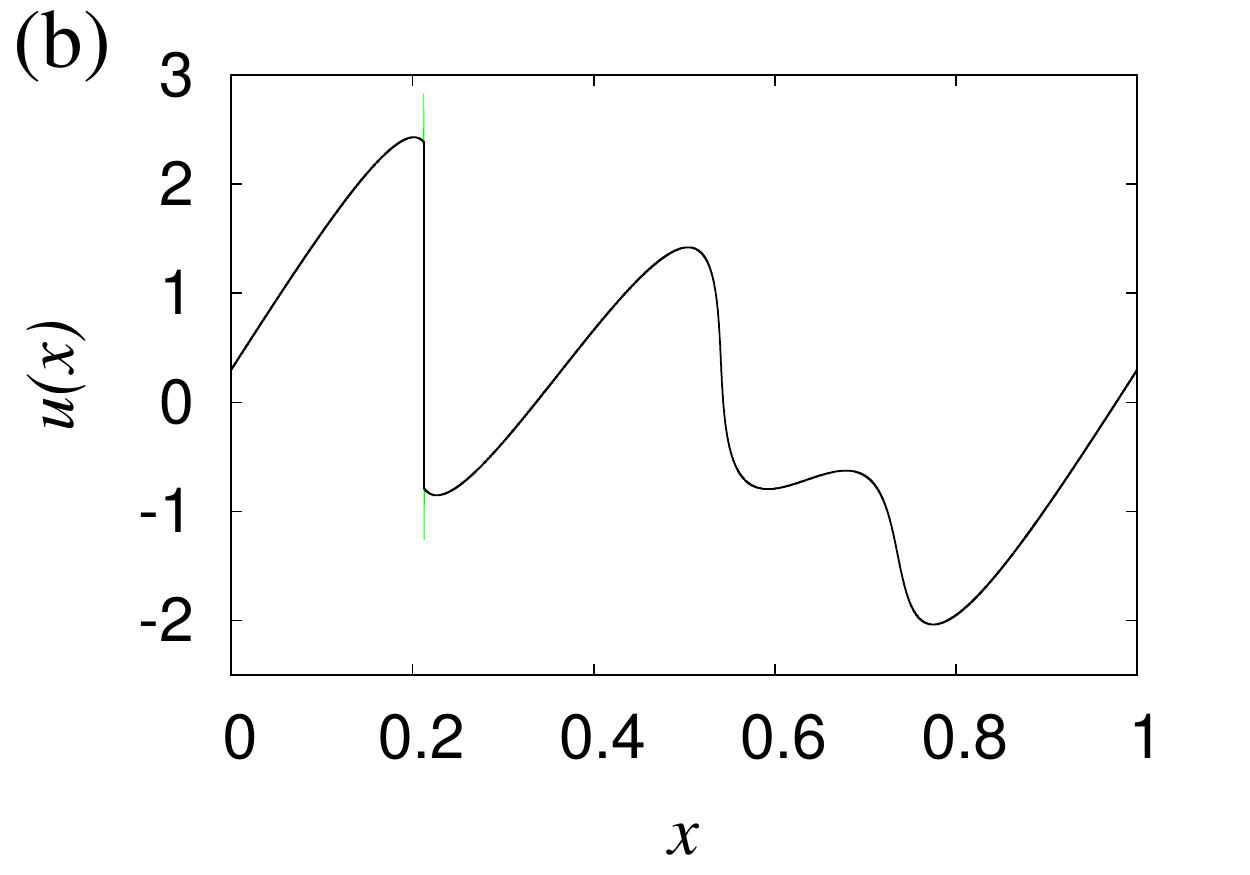}}\hspace{0.15cm}
\subfloat{\label{fig:kings_ref_129}\includegraphics[width=0.32\textwidth, trim=0cm 0cm 0.8cm 0cm, clip=true]{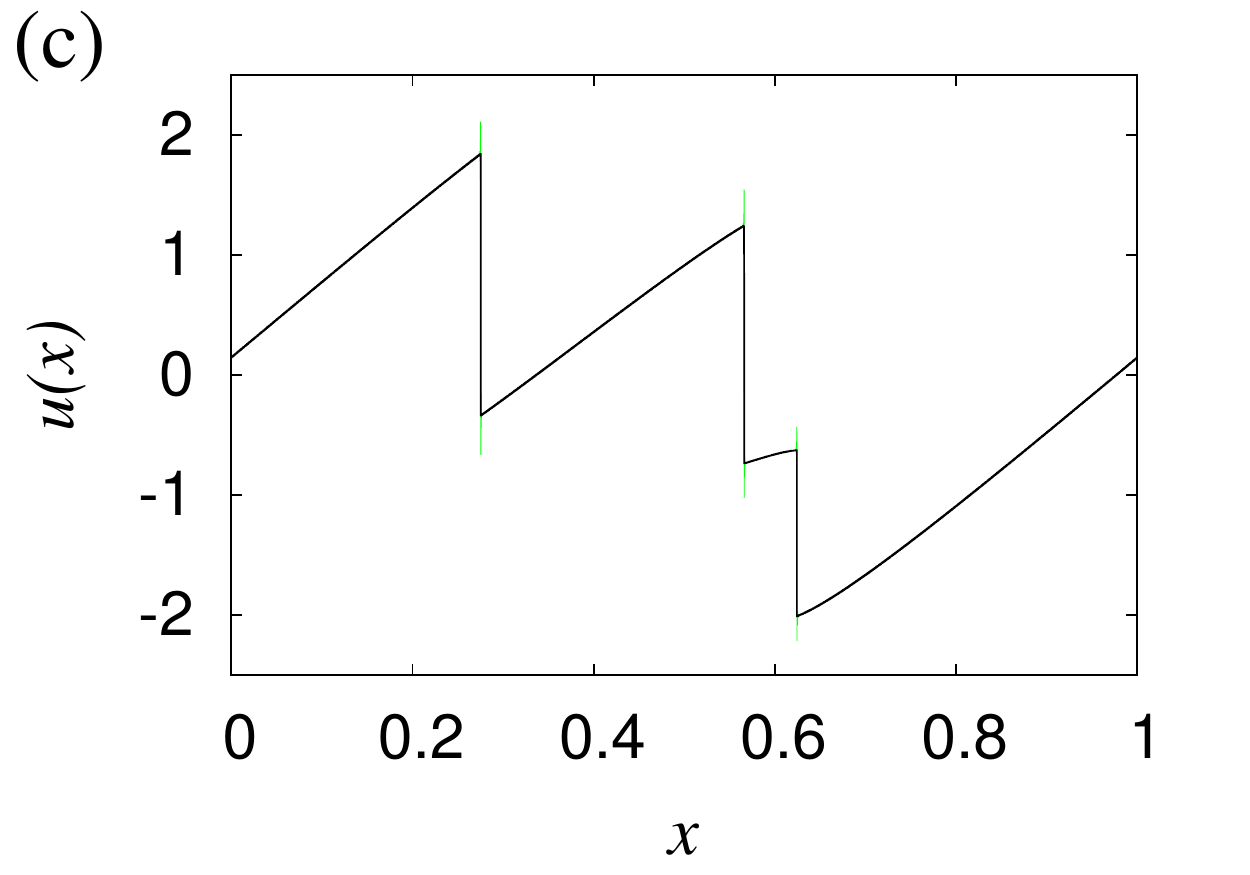}}
\caption{ 
Solutions of the truncated inviscid Burgers equation at $t=0.037$ (a), $t = 0.048$ (b), and $t = 0.129$ (c).
Green (gray): CVS filtered with Kingslets. Black: analytical solution.}
\label{fig:kings_ref}
\end{figure*}

To demonstrate that the whole dynamics of the Burgers equation is preserved by CVS filtering, we plot in Fig.~\ref{fig:kings_ref} the filtered profile along with the analytical solution as a reference, calculated using a Lagrangian map method \cite{Vergassola1994}.
One sees a very good agreement with only small discrepancies at the shocks due to the Gibbs phenomenon.

Overall it appears that this implementation of the CVS filtering method achieves sufficient energy dissipation
at shock locations to keep the numerical solution close to the desired entropy solution.
It would be interesting to understand which element in the algorithm is essential
for this beneficial dissipative effect, but unfortunately there are several competing influences which are difficult to disentangle.
The filtering operation in itself (discarding the incoherent coefficients) is certainly an important
source of dissipation, but it is difficult to quantify a priori
since the Kingslets complex-valued wavelets are not an orthogonal basis, but merely a tight frame (see \cite{NK01}).
Moreover, the alternating projections between the Fourier basis and a wavelet basis, which do not commute which each other,
also introduce some dissipation.
A first step in order to better understand the process by which this filter achieves dissipation
is to move from a wavelet frame to an orthogonal wavelet basis, as we discuss in the next section.


\subsection{Elimination of resonances using real-valued orthogonal wavelets}\label{sec:ROW}

Although the Kingslet frame is well suited to suppress resonances as we have recalled in the previous section, 
it is appealing to be able to use a non-redundant real-valued orthogonal wavelet basis. 
Due to its lack of translation invariance, this kind of basis does not perform well in the context of the algorithm described in the previous section \cite{Nguyenvanyen2009}. 
Following previous work on CVS filtering of the 2D and 3D Navier-Stokes equations \cite{Froehlich1999,Schneider2006,Okamoto2011}, 
we introduce the concept of a safety zone in wavelet space, that is, after computing the coherent coefficients as in the $4^\mathrm{th}$ step of the CVS algorithm, we also keep the neighboring wavelet coefficients in space and in scale. 
The aim is to account for translation of shocks to neighboring positions and generation of finer scale structures from coarser ones. 
Hence, we have to add a step 4b. to the algorithm described in section \ref{sec:kingslet} as follows:

\begin{enumerate}
  \item[4b.]\setcounter{enumi}{4} \emph{Definition of the safety zone in wavelet space}. 
We create an index set $\Lambda$ containing pairs $\lambda = (j,i)$ indexing each coherent wavelet coefficient in scale $j$ and position $i$, kept in step 4. 
We then define an expanded index set $\Lambda_*$ including the neighboring coefficients in position and scale, namely, for each pair $(j,i)$, 
the pairs depicted in Fig.~\ref{fig:safetyzone} \cite{Schneider1996}. Finally, all the coefficients not present in $\Lambda_*$ are set to zero.
\begin{figure}
\setlength{\unitlength}{0.14in} 
\begin{picture}(25,9)  
\put(7.55,0.5){\framebox(14.05,2.8){(j-1,[i/2])}}
\put(0.51,3.3){\framebox(7,2.8){(j,i-1)}}
\put(7.55,3.3){\framebox(7,2.8){(j,i)}}
\put(14.6,3.3){\framebox(7,2.8){(j,i+1)}}
\put(7.55,6.15){\framebox(3.5,2.8){(j+1,2i)}}
\put(11.1,6.15){\framebox(3.45,2.8){}}
\put(11.1,6.7){\makebox(3.45,2.8){(j+1,}}
\put(11.1,6.6){\makebox(3.45,2.8)[b]{2i+1)}}
\end{picture}
\caption{Definition of the safety zone around the point (j,i).}
\label{fig:safetyzone}
\end{figure}
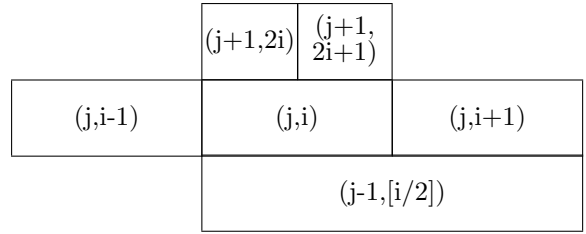
\end{enumerate}
This additional step is able to generate a more stable solution, but the fluctuation level is still high when compared to Kingslets. 
In order to smooth out these fluctuations we need a higher threshold in the CVS filter step of the algorithm, so we choose $q=8$ in equation (\ref{threshold}), changing accordingly the start-up value $T_0$. 

\begin{figure*}
\centering
\subfloat{\label{fig:daub12_safenosafe_067}\includegraphics[width=0.32\textwidth, trim=0cm 0cm 0.8cm 0cm, clip=true]{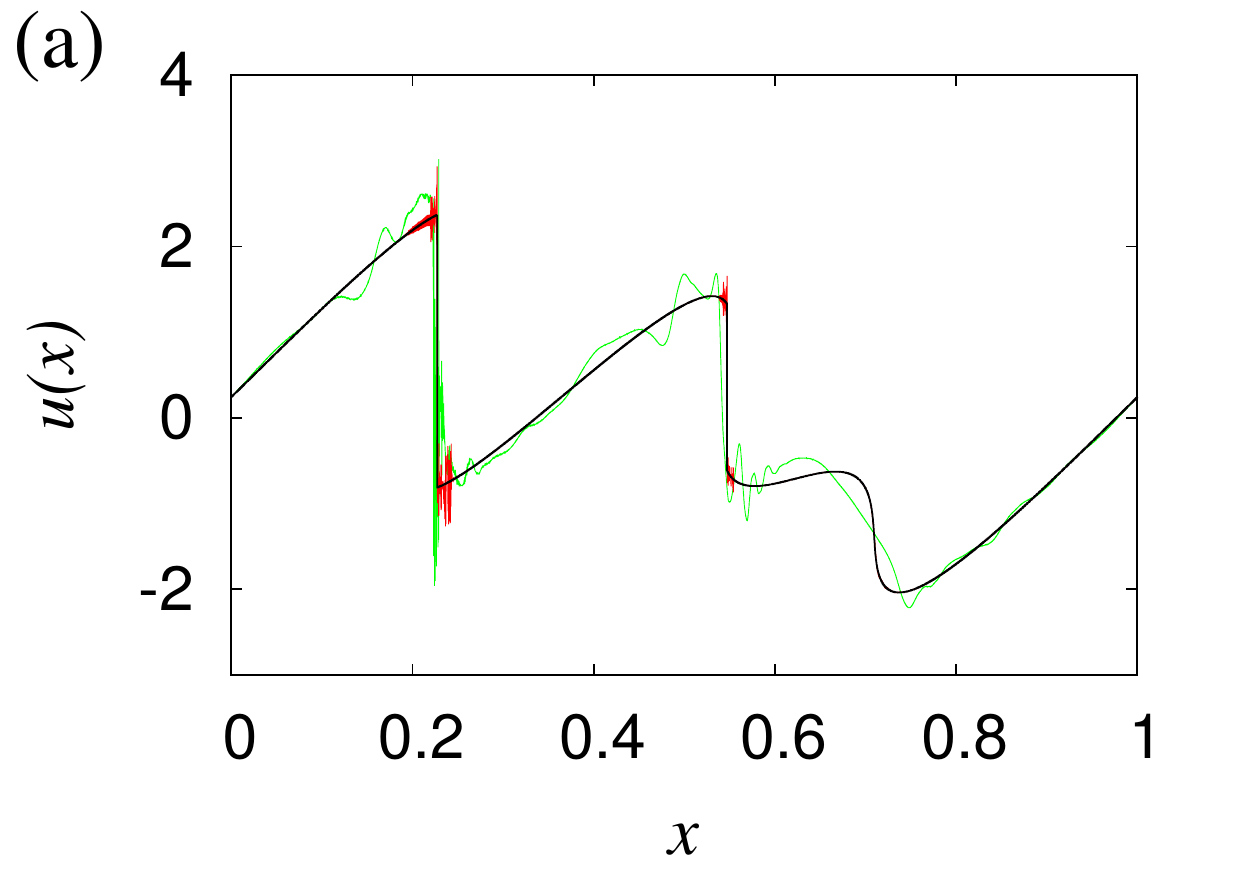}}\hspace{0.15cm}
\subfloat{\label{fig:daub12_safenosafe_143}\includegraphics[width=0.32\textwidth, trim=0cm 0cm 0.8cm 0cm, clip=true]{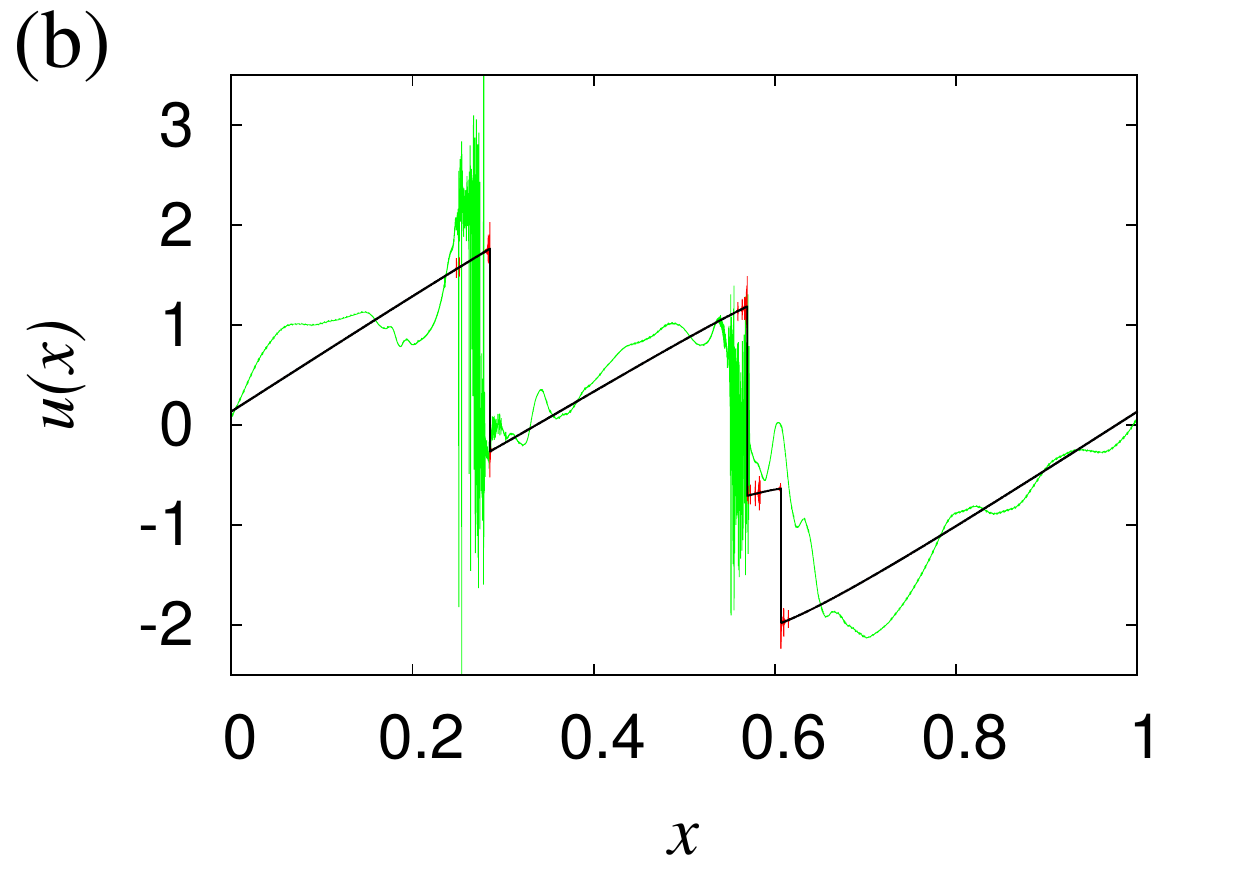}}\hspace{0.15cm}
\subfloat{\label{fig:daub12_safenosafe_200}\includegraphics[width=0.32\textwidth, trim=0cm 0cm 0.8cm 0cm, clip=true]{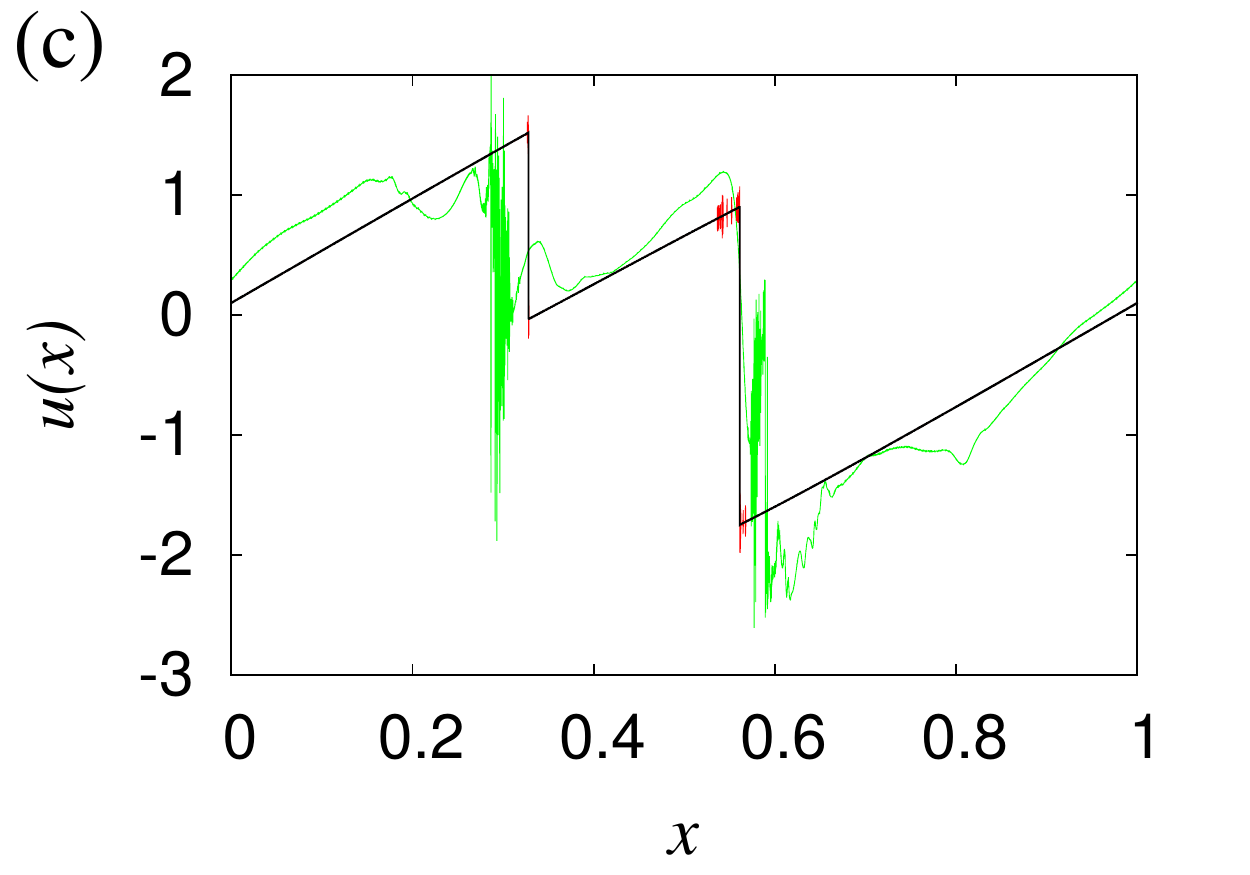}}
\caption{ 
Solutions of the truncated inviscid Burgers equation, CVS-filtered with a periodic Daubechies 12 basis,
at $t=0.067$ (a), $t = 0.143$ (b), and $t = 0.200$ (c).
Green (light gray): no safety zone. Red (dark gray): with safety zone. Black: analytical solution.}
\label{fig:daub12_safenosafe}
\end{figure*}

\begin{figure*}
\centering
\subfloat{\label{fig:spline_safenosafe_067}\includegraphics[width=0.32\textwidth, trim=0cm 0cm 0.8cm 0cm, clip=true]{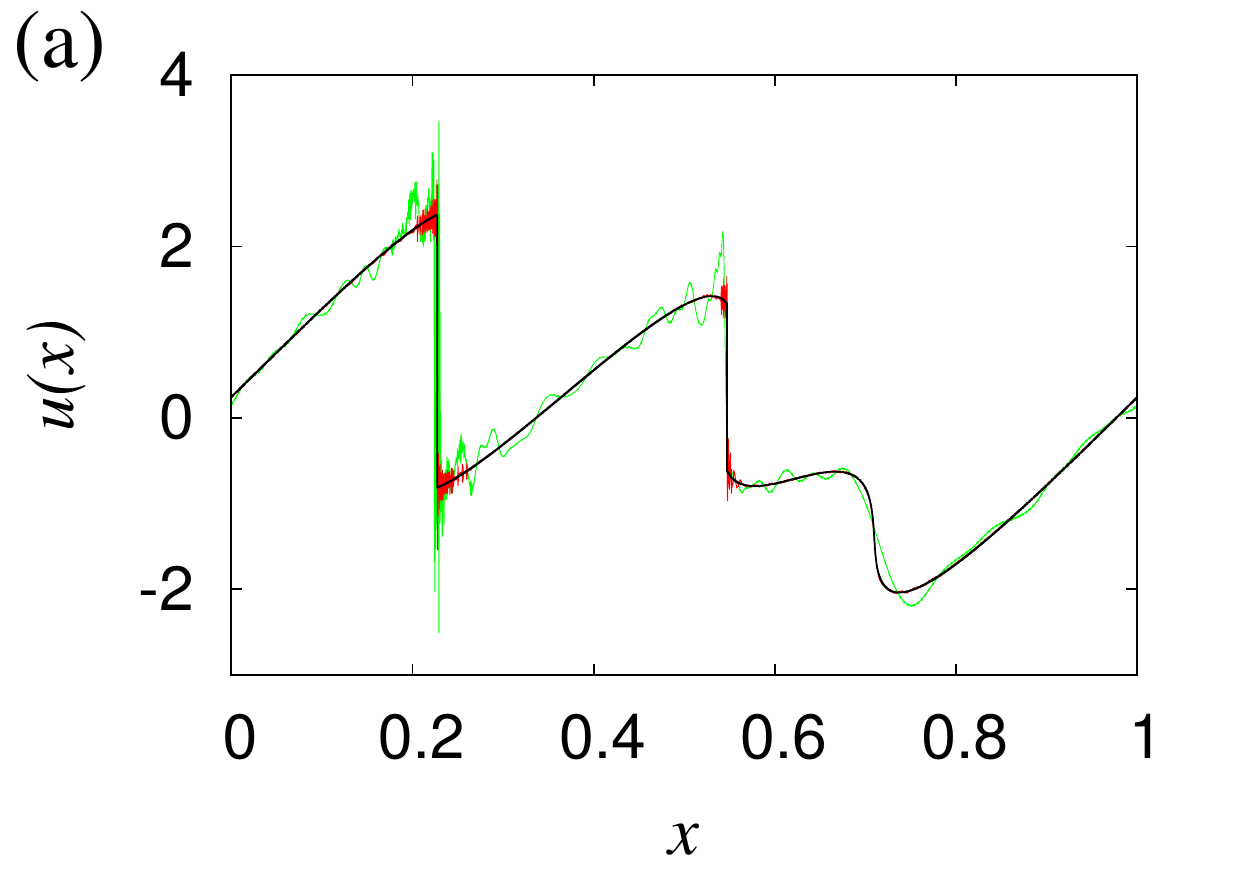}}\hspace{0.15cm}
\subfloat{\label{fig:spline_safenosafe_143}\includegraphics[width=0.32\textwidth, trim=0cm 0cm 0.8cm 0cm, clip=true]{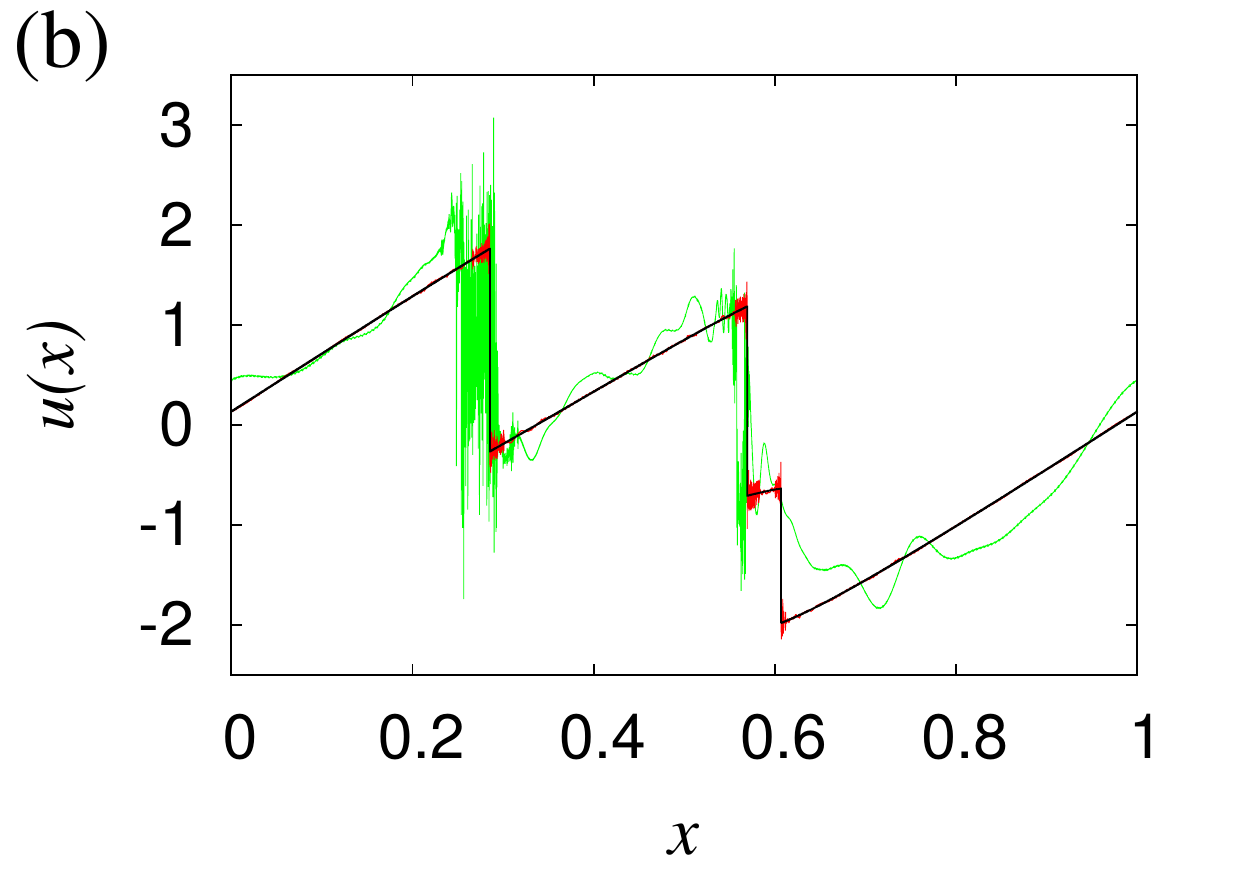}}\hspace{0.15cm}
\subfloat{\label{fig:spline_safenosafe_200}\includegraphics[width=0.32\textwidth, trim=0cm 0cm 0.8cm 0cm, clip=true]{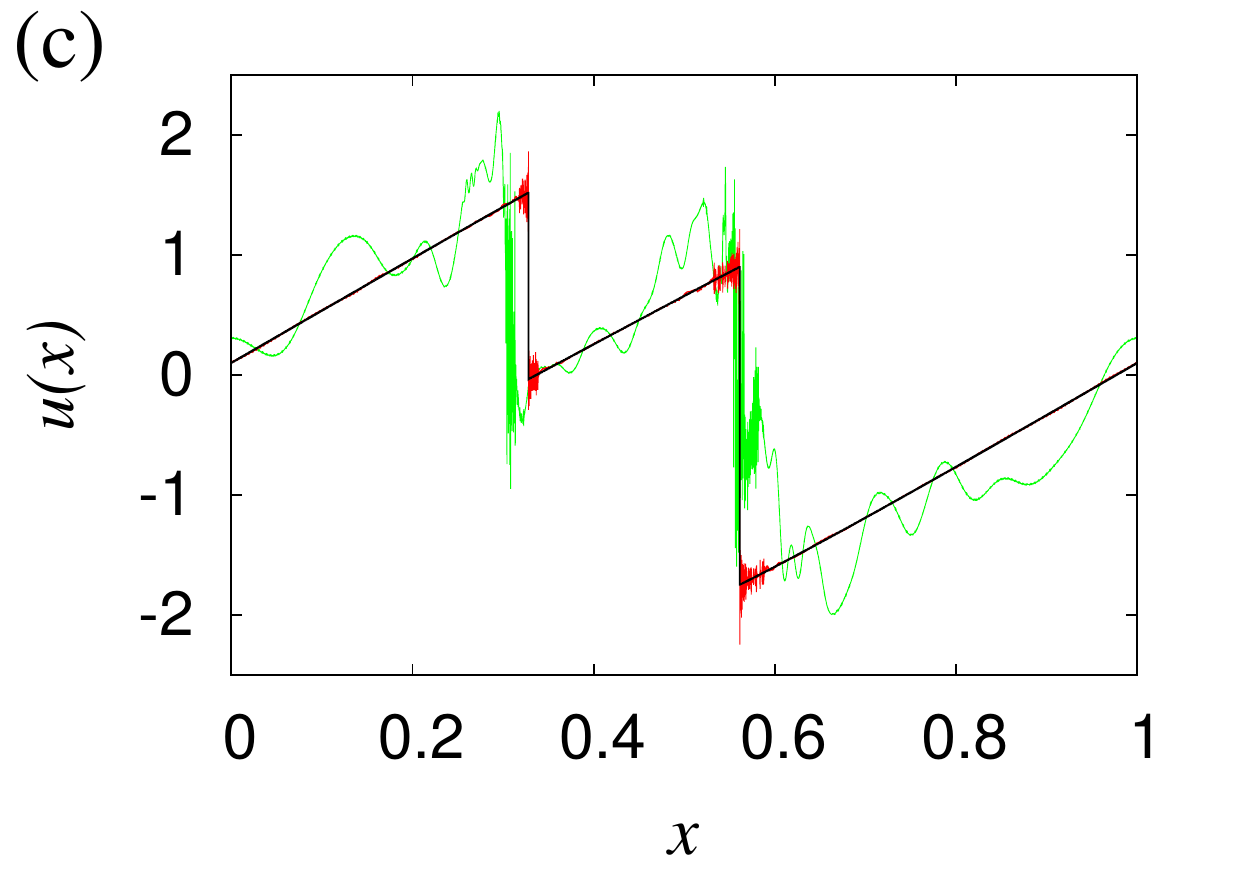}}
\caption{ 
Solutions of the truncated inviscid Burgers equation, CVS-filtered with a periodic Spline 6 basis,
at $t=0.067$ (a), $t = 0.143$ (b), and $t = 0.200$ (c).
Green (light gray): no safety zone. Red (dark gray): with safety zone. Black: analytical solution.}
\label{fig:spline_safenosafe}
\end{figure*}

As examples we employ two different wavelet bases that are widely available in numerical analysis packages, 
the Daubechies 12 wavelet, which has compact support, and the Spline 6 wavelet, which has an exponential decay \cite{Daubechies1992}. 
If we simply apply the CVS filtering procedure from section \ref{sec:kingslet} with these bases, 
the solution becomes very oscillating as soon as the resonances appear and we end up with poor results. 
But once the safety zone in wavelet space is implemented as described above, the dynamics is properly preserved. 
In Figs.~\ref{fig:daub12_safenosafe} (Daubechies 12) and \ref{fig:spline_safenosafe} (Spline 6) we see the significant improvement in the filtering capability of the code, comparing the cases with and without safety zone along with the analytical solution.

\begin{figure}
\centering
\includegraphics[width=0.45\textwidth]{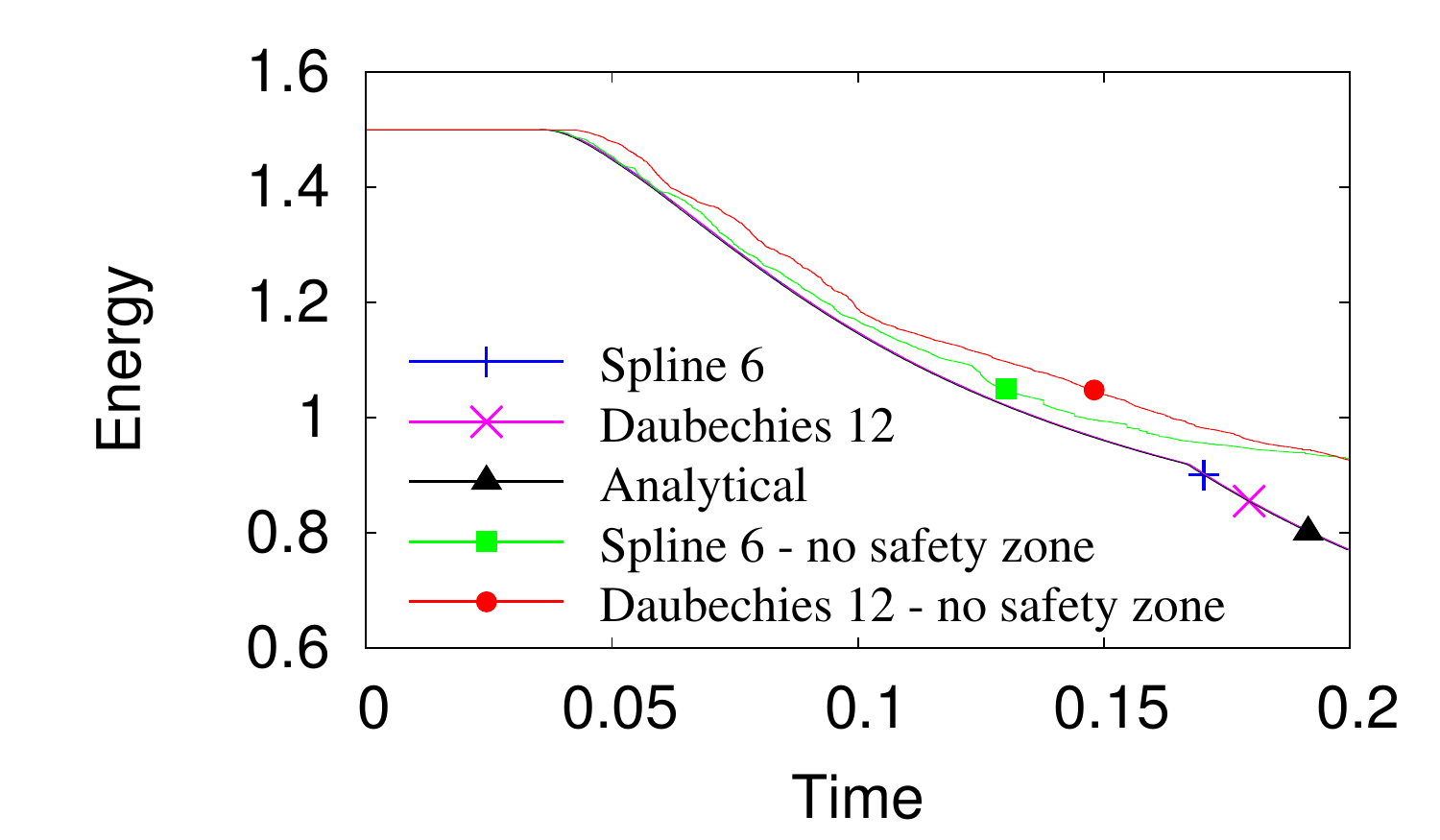}
\caption{ Time evolution of energy. Energies of the CVS filtered solutions with safety zone collapse to the analytical energy evolution.}
\label{fig:energy}
\end{figure}

The naturally oscillating character of real-valued wavelets and their lack of translation invariance still plays a role generating small perturbations (that get worse next to regions affected by the Gibbs phenomenon). 
But while the dynamics is lost when there is no safety zone, with huge oscillations corrupting the phase coherence of the shocks, after the introduction of the safety zone it is very well preserved. 
Considering the time evolution of energy (Fig.~\ref{fig:energy}), it appears that in absence of a safety zone, not all the necessary energy is dissipated.
This could be an explanation for the poor performance of the filtering scheme in that case.

\begin{figure*}
\centering
\subfloat{\label{fig:errors1}\includegraphics[width=0.4\textwidth]{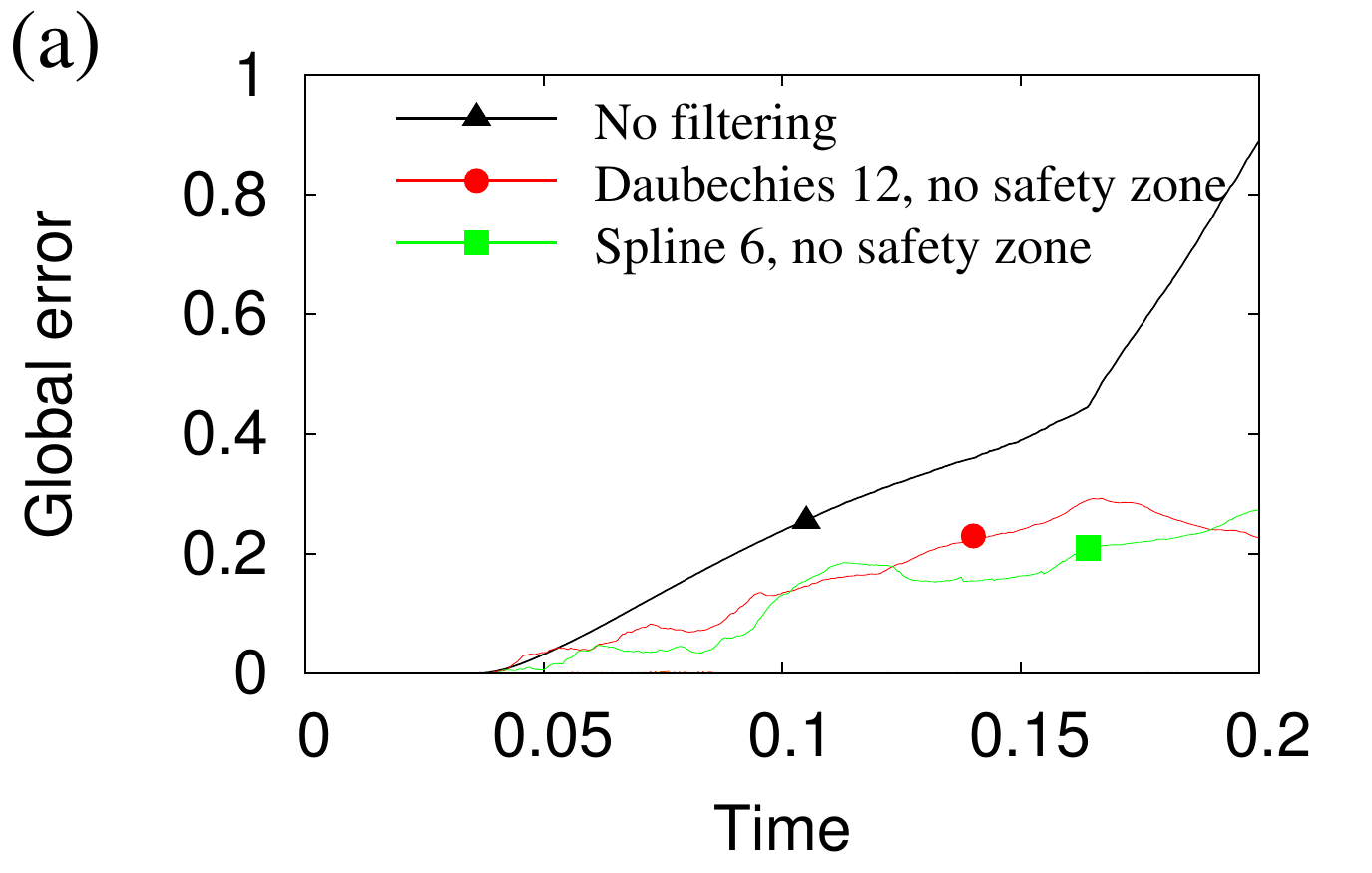}} \hspace{1cm}
\subfloat{\label{fig:errors2}\includegraphics[width=0.42\textwidth]{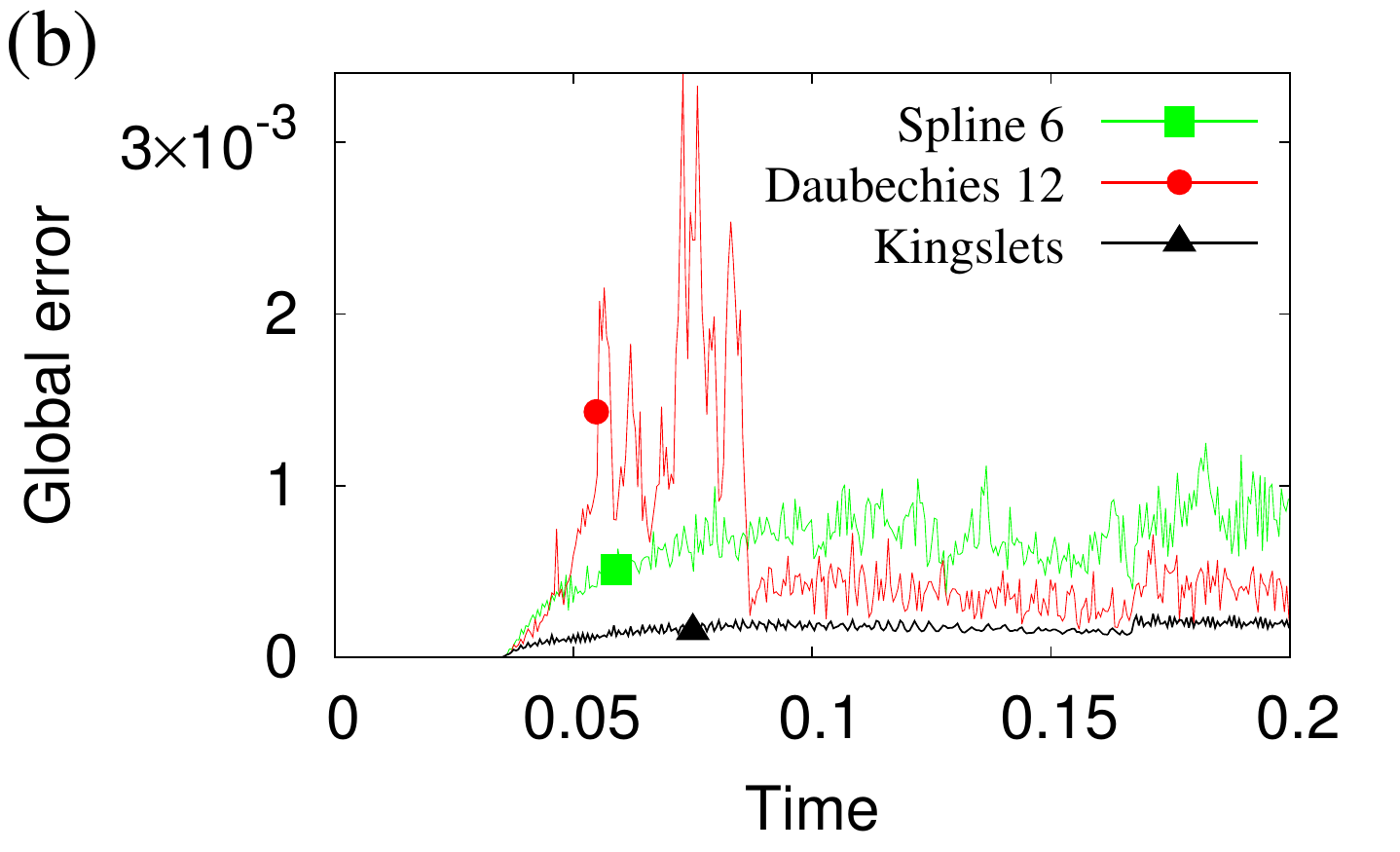}}
\caption{ (a) Time evolution of the global energy error $\varepsilon$. (b) Rescaling of (a). 
}
\end{figure*}

To give a quantitative aspect to the idea of ``good filtering'' we consider the global energy error estimate 
\eqn{\varepsilon = \frac{\int_0^1 \left[v(x) - v_{\mathrm{ref}}(x)\right]^2 dx}{\int_0^1 v_{\mathrm{ref}}(x)^2 dx},}
where $v_{\mathrm{ref}}$ is the reference analytical solution. 
This allows us not only to evaluate how close to the reference the CVS-filtered solutions are, but also to compare the efficiencies of different wavelet bases. 
In Fig.~\ref{fig:errors1} we plot the time evolution of $\varepsilon$ for all runs.
\begin{figure}
\centering
\includegraphics[width=0.4\textwidth]{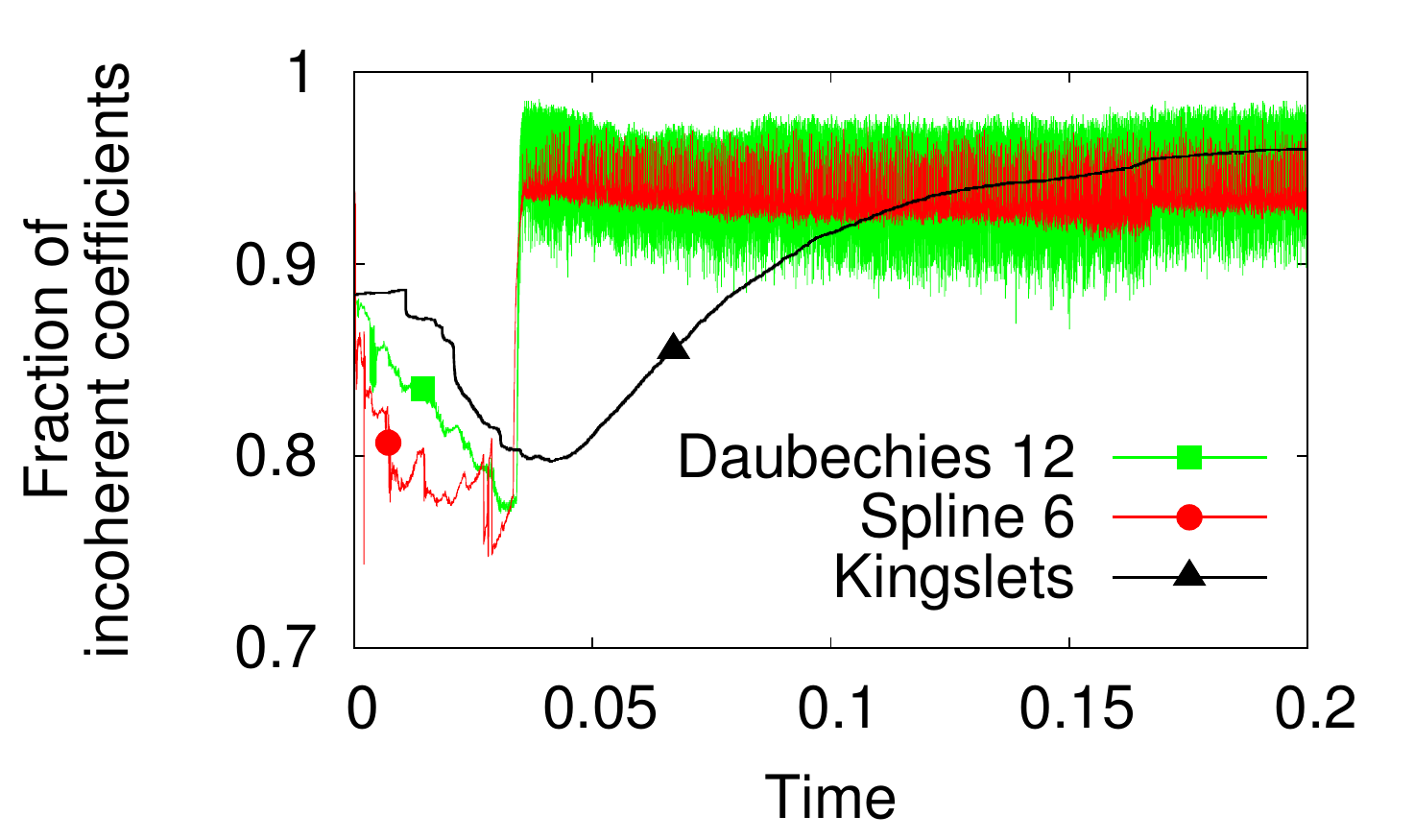}
\caption{\label{fig:incoh} Evolution of the fraction of incoherent wavelet coefficients.}
\end{figure}
For the unfiltered Galerkin-truncated solution, the error grows very fast as soon as resonances appear. 
The growth is slower for the CVS solutions without safety zone, but the solutions are still eventually destroyed. 
Due to their much smaller values, the error estimates for the Kingslets and for the real-valued wavelets with safety zone are barely seen in this plot. 
So in Fig.~\ref{fig:errors2} we change scales to find that they are of the order $10^{-3}$ and stabilize once the influence of the resonances has been damped. 
We see that the errors of the real-valued orthogonal wavelets stabilize very close to the Kingslets value. 
This makes their use attractive, a fact even more reinforced when we compare the level of compression along the time evolution (Fig.~\ref{fig:incoh}), \emph{i.e.}, the percentage of discarded coefficients each time step.

Indeed, during a large fraction of the evolution, Kingslet-based CVS filtering keeps many more coefficients than its counterparts based on orthogonal wavelets.
The level of compression tends to stabilize at a slightly smaller value than the average of the other cases, 
but since the Kingslets frame has twice as many coefficients as real-valued orthogonal wavelet bases, this result shows the strong potential of the latter
for the development of fully adaptive methods, provided a safety zone is implemented. 


\section{2D Euler equation}

The emergence of resonances in Galerkin-truncated solutions of the 2D Euler equation was also shown in \cite{Ray2011}. 
The fact that the CVS solutions, filtered with a 2D version of the Kingslets, are similar to the ones obtained from 2D Navier-Stokes with small viscosity \cite{Nguyenvanyen2009} suggests that CVS might be suitable to filter the resonances in this case as well.
Therefore, in the same spirit as in section \ref{sec:kingslet}, we apply the CVS method using Kingslets to the same initial condition used in the 2D example of \cite{Ray2011}:
\begin{equation}\label{eq:euler_initial_condition}
\widehat{\omega}_\kk = \frac{2\vert k \vert^{7/2}}{N_k} e^{-k^2/4 + i\theta_\kk},
\end{equation}
where $\theta_\kk$ is a realization of a random variable uniformly distributed in $[0,2\pi]$, $k$
is the integer part of $\vert \kk \vert$, and $N_k$ is the number of distinct vectors $\kk$ such that $ k \leq \vert \kk \vert < k+1$.
The particular realization used in \cite{Ray2011} as well as here can be retrieved online \footnote{\url{http://www.kyoryu.scphys.kyoto-u.ac.jp/\%7Etakeshi/populated}}.
The 2D Galerkin-truncated Euler equations are solved using a fully dealiased pseudo-spectral method at resolution $N^2 = 1024^2$ with a low storage third order Runge-Kutta scheme for time discretization.
The time step is adjusted dynamically to satisfy the CFL stability criterion.
For more details on the numerical method, we refer the reader to \cite{Nguyenvanyen2009}.

\begin{figure*}
\centering
\includegraphics[width=0.4\textwidth]{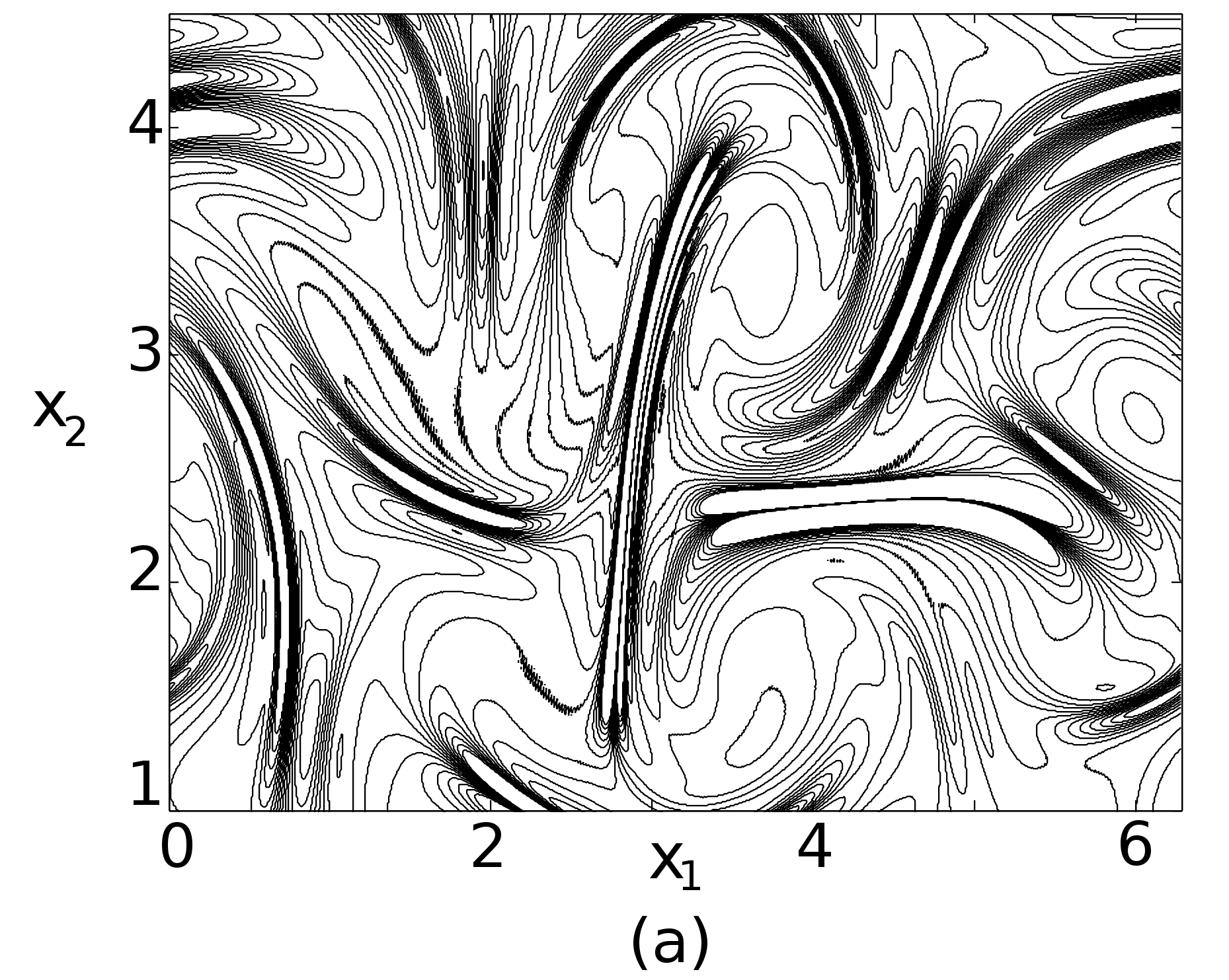} \hspace{1cm} 
\includegraphics[width=0.4\textwidth]{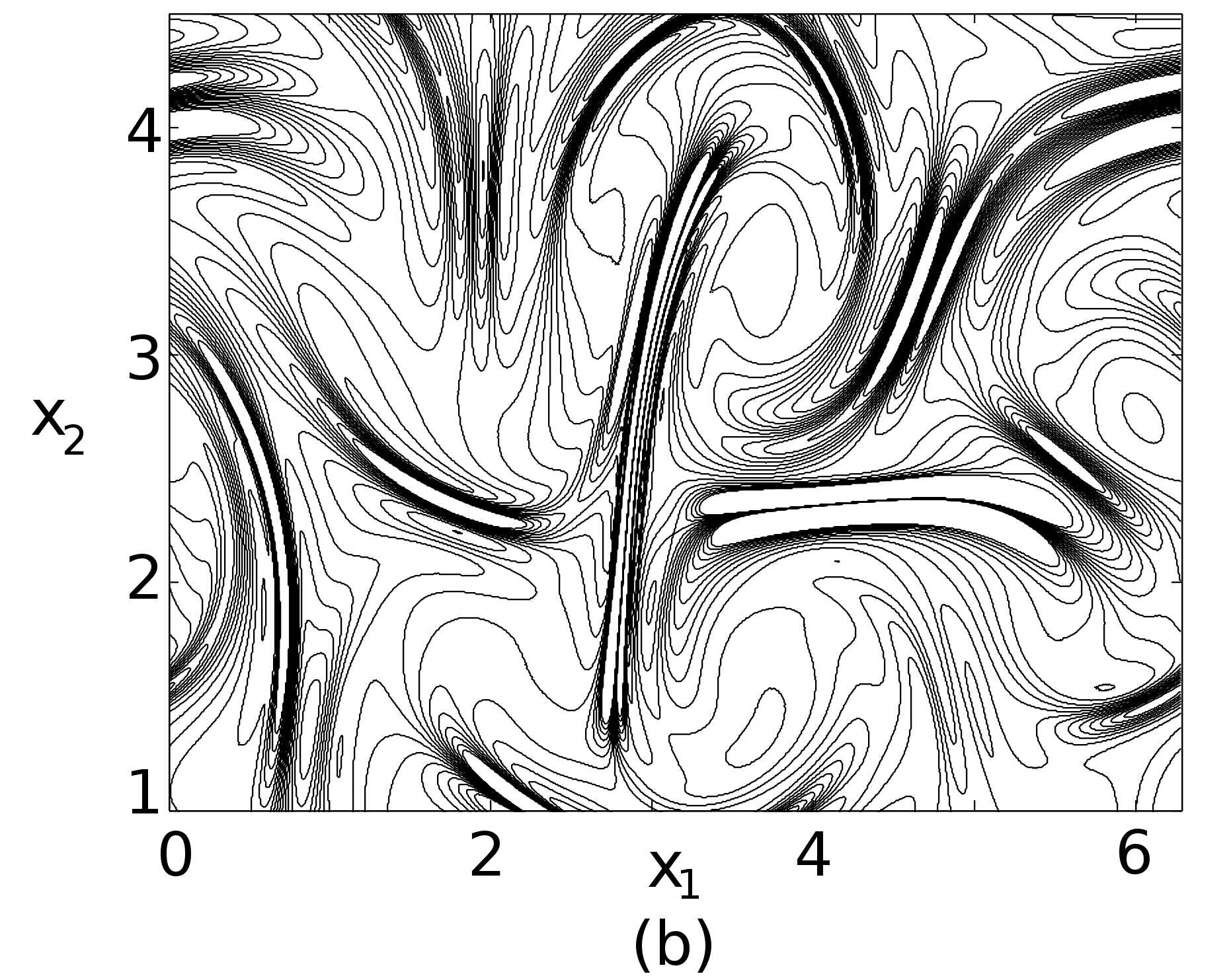} 
\caption{Contours of the Laplacian of vorticity at $t=0.71$ (from $-200$ to $200$, increments of $25$). (a) Galerkin-truncated. (b) CVS.
}
\label{fig:euler}
\end{figure*}

\begin{figure*}
\centering
\includegraphics[width=0.4\textwidth]{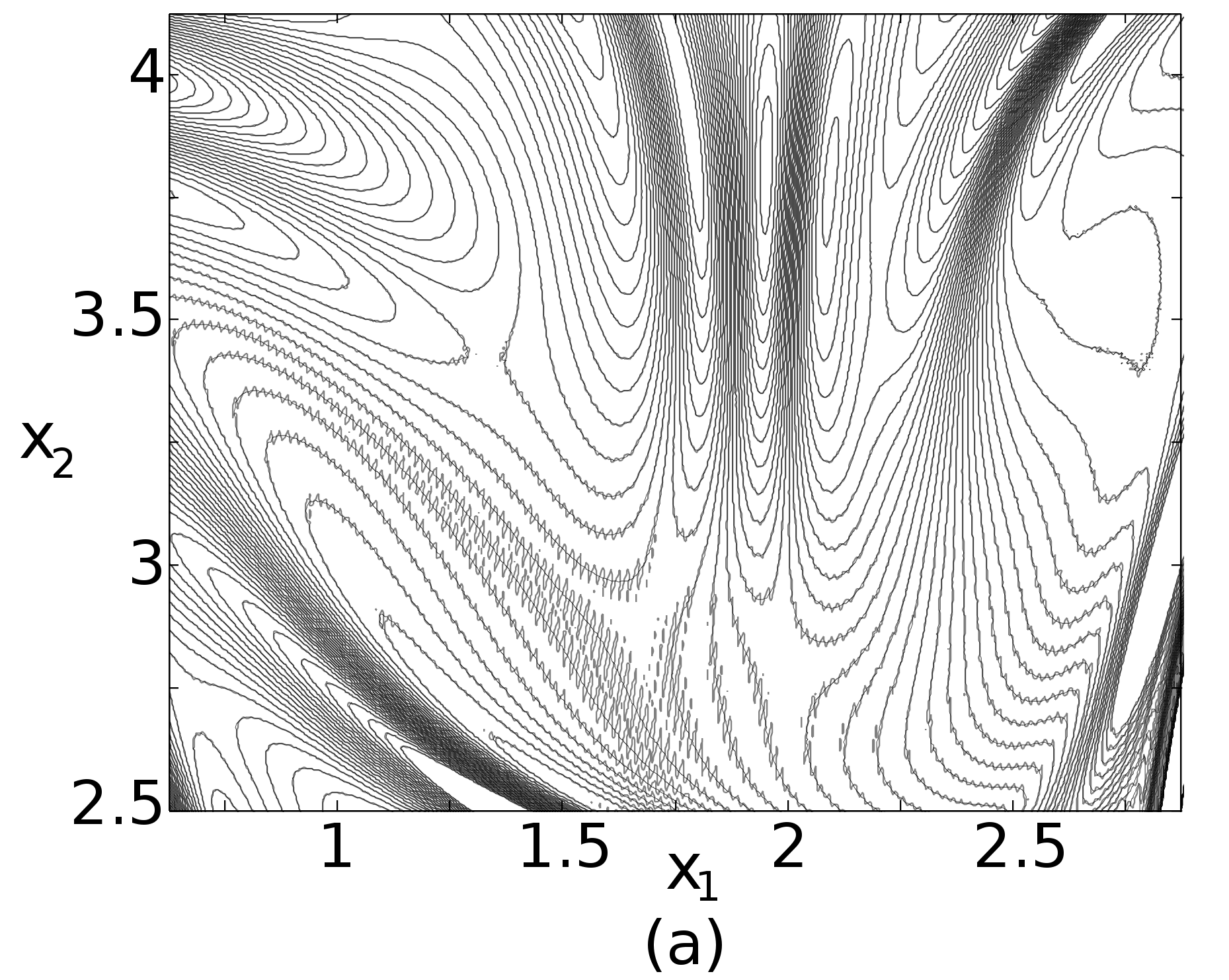}  \hspace{1cm} 
\includegraphics[width=0.4\textwidth]{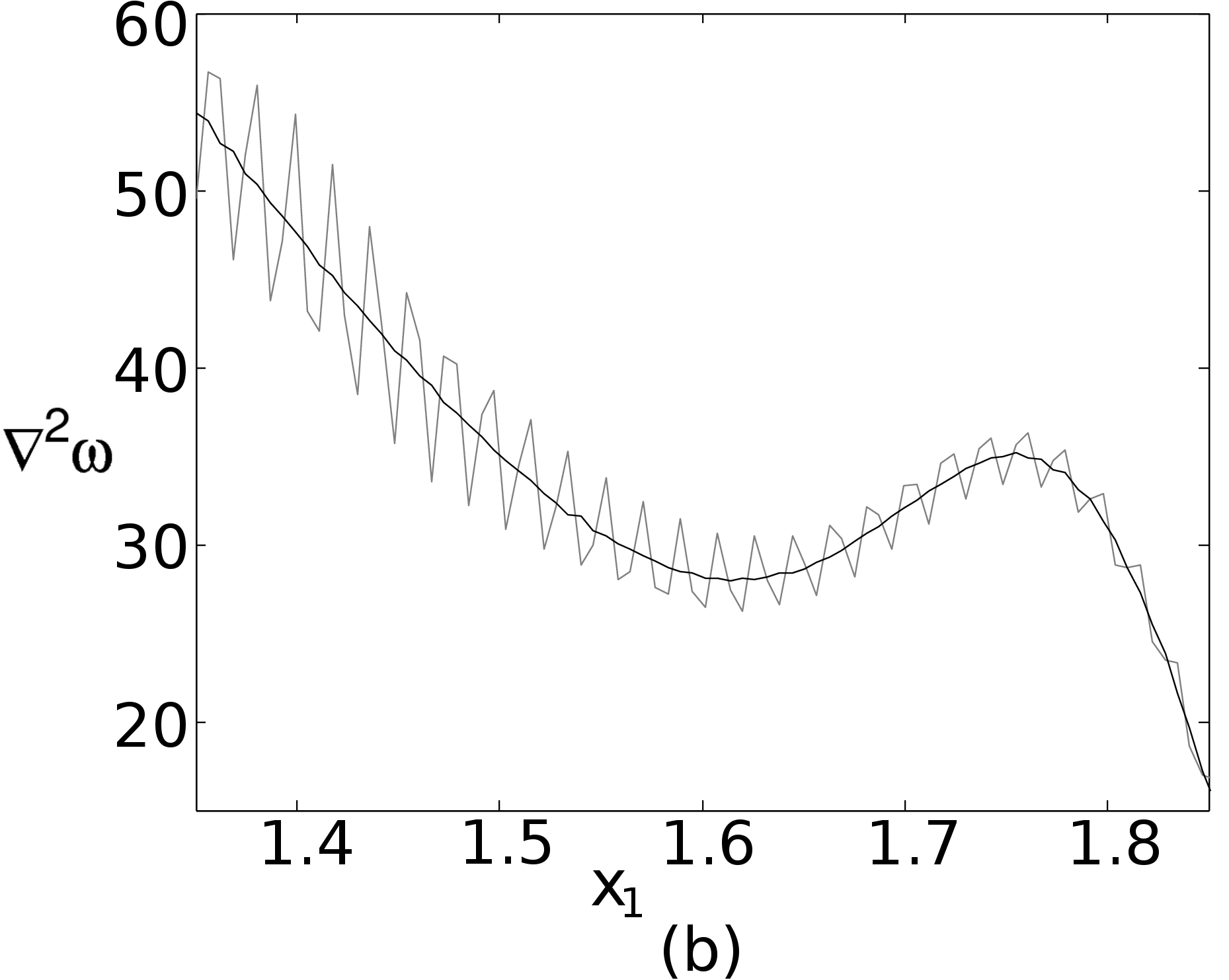}
\caption{
(a) Zoomed contours of the Laplacian of vorticity at $t=0.71$. Gray: Galerkin-truncated. Black: CVS.
(b) Laplacian of vorticity at $t=0.71$ along a segment near to $x_2=3$. Gray: Galerkin-truncated. Black: CVS.
}
\label{fig:comp_euler_zoom}
\label{fig:comp_euler_lap}
\end{figure*}

\begin{figure}
\centering
\includegraphics[width=0.42\textwidth]{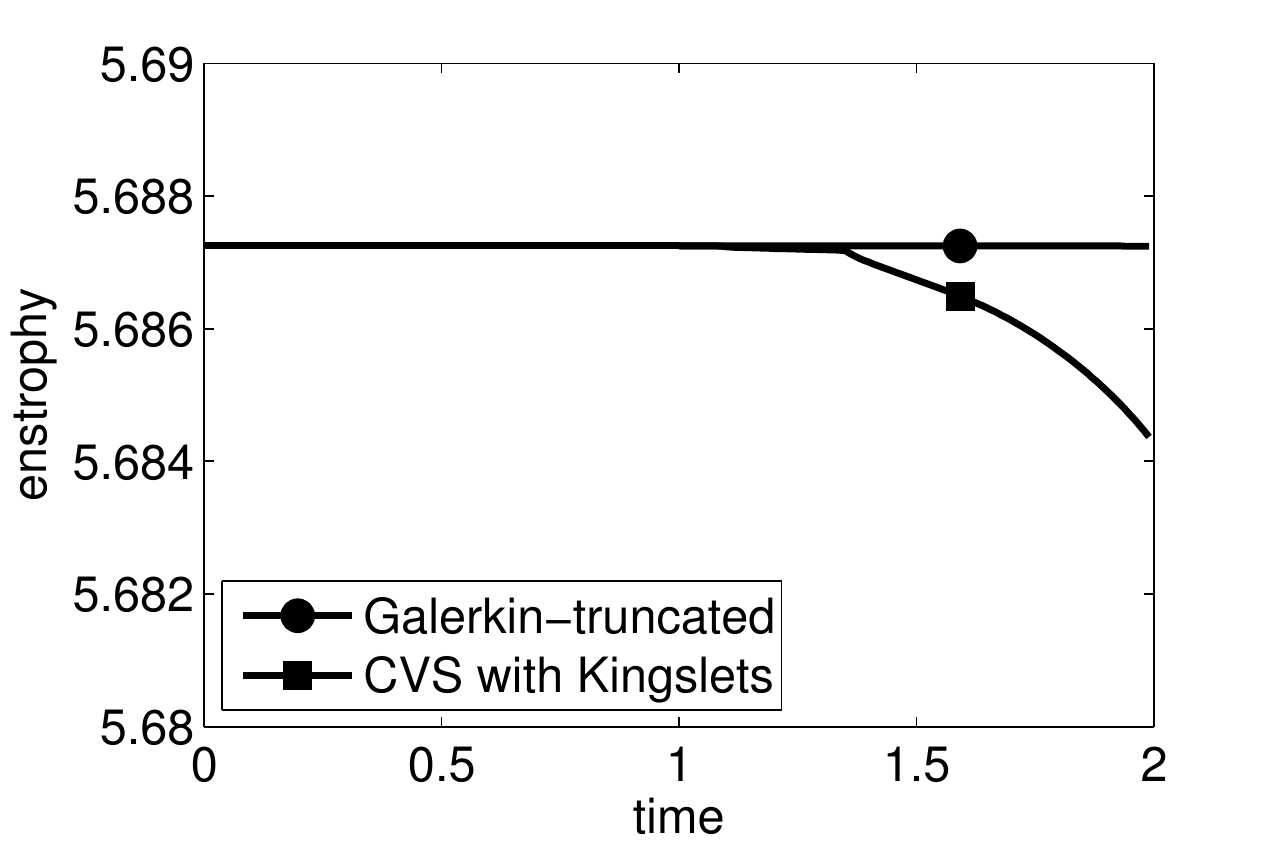}
\caption{Time evolution of enstrophy for the Galerkin-truncated and CVS-filtered Euler equations.}
\label{fig:comp_euler_enstrophy}
\end{figure}

In contrast to the Burgers case previously presented, we do not have here an analytical solution to compare with,
and make an error estimate, but a visual qualitative comparison will be sufficient to check if CVS filters out the resonances while preserving the dynamics. 
Resonances are well exhibited in plots of the Laplacian of vorticity, so, following \cite{Ray2011}, we show contours of this quantity at $t=0.71$.
Figure~\ref{fig:euler} shows the contours for the whole domain and we can easily see that CVS solutions do not show the resonances but keep the same general aspect.
A more precise comparison can be made from Fig.~\ref{fig:comp_euler_zoom}, where the contours of both cases at $t=0.71$, zoomed-in around a region of intense resonance, are plotted together (left panel), as well as a cut as a function of $x_1$ along a segment near $x_2=3$ (right panel).
One sees very well how the resonances are suppressed and how the profiles are strikingly similar, indicating that the filter is able to maintain the physical aspects of the solutions.
Finally, the dissipative character of the CVS filter is confirmed when considering the time evolution of the enstrophy $Z = \frac{1}{2} \int \omega^2$,
as shown in Fig.~\ref{fig:comp_euler_enstrophy}.
%


\section{Conclusion}

The continuous wavelet transform allowed to get further insight into the scale-space dynamics of resonance phenomena in Galerkin truncated inviscid equations. 
We showed that oscillations appear in a non-local fashion as soon as a shock affects the cut-off scale, and that the resonant points and the shock act as sources of perturbations at the cut-off scale. 
We could also see that despite the fact that the resonances first appear at small scales, 
large scale structures develop at the resonant points and are stretched into smaller scales until they reach the cut-off and start acting as new sources of truncation waves. These new perturbations spread and reach the shocks, leading to energy equipartition.

For the 1D inviscid Burgers equation, the results presented here confirm that the CVS filtering method we have previously proposed in [5], using a dual-tree complex wavelet frame (Kingslets), is well suited for eliminating all spurious oscillations present in the Galerkin-truncated solution as reported in \cite{Ray2011}. 
The resonances, which are not due to the dynamics of the original equation but rather to its discretization by a Galerkin method, are completely suppressed in this approach. 
Their `incoherent' character in relation to the system evolution is established.
In order to better understand the dissipative process characteristic of CVS filtering, we have sought to replace Kingslets by standard real-valued orthogonal wavelets.
We have obtained satisfactory results under the condition that the coefficients which are adjacent to those whose moduli are above the threshold value are preserved.
Such a safety zone is only necessary with orthogonal wavelets, to compensate for their lack of translation invariance, as originally introduced for CVS filtering of the 2D and 3D Navier-Stokes equations \cite{Froehlich1999,Schneider2006,Okamoto2011}.

For the 2D Euler equation we have shown that CVS filtering with Kingslets is also capable of filtering the resonances without perturbing the dynamics. 
The filtered solutions match the unfiltered ones but for the non-physical oscillations which are eliminated.
The authors of \cite{Ray2011} asserted that many features of the resonance phenomena were also observed in the 3D Galerkin-truncated Euler equations, though these results have not been reported yet. 
It is an interesting perspective to test if in this case CVS filtering is still able to eliminate the resonances.

A limitation of the approach presented here is that the solution is transformed back and forth at each timestep between the wavelet and the Fourier truncations,  
which do not commute with each other. 
These alternating projections are likely to introduce a weak dissipation in addition to the filtering operation \textit{per se}.
Therefore from the present results it cannot yet be determined whether the observed elimination of resonances could be achieved solely with wavelet filtering,
or whether the interleaved truncations in Fourier space play a crucial role.
This question could be answered by applying the filtering method to the Wavelet-Galerkin truncation of the equations,
instead of the Fourier-Galerkin truncation that was considered here, offering an appealing perspective for future work.

\section*{Acknowledgments}

RMP thanks the Brazilian National Scientific and Technological Research Council (CNPq) for support.
RNVY thanks the ANR Geofluids and the Humboldt foundation for supporting this research through post-doctoral fellowships.
RNVY, MF and KS acknowledge support from the contract Euratom-FR-FCM n°2TT.FR.1215 and from the PEPS program of CNRS-INSMI.
They are also grateful to M. Domingues and O. Mendes for their kind hospitality in Brazil while revising this paper.


\bibliography{biblio}
\bibliographystyle{plain}

\end{document}